\documentclass[fleqn,usenatbib]{mnras}

\usepackage{graphicx}
\usepackage{hyperref}
\usepackage{tabularx}
\usepackage{mathtools}
\usepackage{amsmath,amssymb}
\usepackage{chngcntr}
\usepackage[usenames,dvipsnames]{color}
\DeclareUnicodeCharacter{2212}{-}

\DeclareRobustCommand{\VAN}[3]{#2}
\let\VANthebibliography\thebibliography
\def\thebibliography{\DeclareRobustCommand{\VAN}[3]{##3}\VANthebibliography}

\usepackage{graphicx}	% Including figure files
\usepackage{amsmath}	% Advanced maths commands
\usepackage{amssymb}	% Extra maths symbols
\usepackage{multicol}        % Multi-column entries in tables
\usepackage{bm}		% Bold maths symbols, including upright Greek
\usepackage{pdflscape}	% Landscape pages
\usepackage[flushleft]{threeparttable} %to create table notes
\usepackage{subcaption} %for multiple panels in one figure

\newcommand{\fcgs}{\mathrm{erg\,s^{-1}\,cm^{-2}}}

\title[Quasar Circumgalactic Medium in H$\alpha$ and Ly$\alpha$ Emission]{Characterizing the Circumgalactic Medium of Quasars at z$\sim$2.2 through H$\alpha$ and Ly$\alpha$ Emission}

\author[V. Langen et al.]{Vivienne Langen$^{1,2,4}$\thanks{E-mail: vivienne.langen@l2it.in2p3.fr}, 
Sebastiano Cantalupo$^{3,2}$\thanks{E-mail: sebastiano.cantalupo@unimib.it}, 
Charles C. Steidel$^{4}$, 
Yuguang Chen$^{4,5}$, \newauthor 
Gabriele Pezzulli$^{6,2}$ and
Sofia G. Gallego$^{4,2}$\\
$^{1}${L2IT, CNRS, Université Paul Sabatier, 118 Route the Narbonne, 31062, Toulouse, France} \\
$^{2}${Department of Physics, ETH Zurich, Wolfgang-Pauli-Strasse 27, 8093, Zurich, Switzerland} \\
$^{3}${Department of Physics, University of Milan Bicocca, Piazza della Scienza 3, 20126, Milano, Italy} \\
$^{4}${Cahill Center for Astrophysics, California Institute of Technology, 91125, Pasadena, CA, USA} \\
$^{5}${University of California, Davis, 1 Shields Ave, Davis, CA, 95616, US} \\
$^{6}${Kapteyn Astronomical Institute, University of Groningen,Landleven 12, 9747 AD Groningen, The Netherlands}
\\
}

\date{Accepted 2022 November 2. Received 2022 November 2; in original form 2022 April 11}

\pubyear{2022}

\begin{document}
\label{firstpage}
\pagerange{\pageref{firstpage}--\pageref{lastpage}}
\maketitle

\begin{abstract}
The discovery of giant quasar Ly$\alpha$ nebulae at $z>2$ has opened up the possibility to directly study in emission the Circumgalactic and Intergalactic Medium (CGM/IGM). However, the resonant nature of the Ly$\alpha$ line and its different emission mechanisms hamper the ability to constrain both the kinematics and physical properties of the CGM/IGM.
 Here, we present results of a pilot project aiming at the detection of CGM H$\alpha$ emission, a line which does not suffer from these limitations. To this end, we first used KCWI to detect Ly$\alpha$ emission around three bright quasars with $2.25<z<2.27$, a range which is free from bright IR sky lines for H$\alpha$, and then selected the most extended nebula for H$\alpha$ follow-up with MOSFIRE. Within the MOSFIRE slit, we detected H$\alpha$ emission extending up to 20 physical kpc with a total H$\alpha$ flux of F$_{\tiny \textrm{H}\alpha}$=(9.5 $\pm$ 0.9) $\times$ 10$^{-18}\fcgs$. Considering the Ly$\alpha$ flux in the same region, we found F$_{\tiny \textrm{Ly}\alpha}$/F$_{\tiny \textrm{H}\alpha}$=3.7 $\pm$ 0.3 consistent with that obtained for the Slug Nebula at z$=2.275$ and with recombination radiation. This implies high densities or a very broad density distribution within the CGM of high-redshift quasars. Moreover, the H$\alpha$ line profile suggests the presence of multiple emitting components overlapping along our line-of-sight and relatively quiescent kinematics, which seems incompatible with either quasar outflows capable of escaping the potential well of the host halo or disk-like rotation in a massive halo ($>10^{12}$M$_{\odot}$).

\end{abstract}


%%%%%%%%%%%%%%%%%%%%%%%%%%%%%%%%%%%%%%%%%%%%%%%%%%

%%%%%%%%%%%%%%%%% BODY OF PAPER %%%%%%%%%%%%%%%%%%

\section{Introduction}
\label{sec:intro}

Current theories of galaxy formation and evolution predict that galaxies should be fueled by gas accretion from their surrounding medium, namely the Circumgalactic or Intergalactic Medium (CGM or IGM). The detailed physical properties of the gas in terms of, e.g., density, temperature and angular momentum, are thus fundamental variables for our understanding of structure formation in the universe. 
Some cosmological simulations \citep{bond1996filaments, fukugita1998cosmic, birnboim2003virial, kerevs2005galaxies, dekel2006galaxy} suggest that between redshift 1.5 $<$ $z$ $<$ 4 most of the accreting material should be relatively cold (T$\sim$10$^{4}$K) and  confined to filaments penetrating the hotter halos \citep{dekel2009cold}. It is however still debated whether such cold filaments could survive the hydrodynamical interaction with the hot halo gas \citep[e.g.][]{mandelker2017giant, vossberg2019density}, or being instead disrupted into a series of smaller and denser structures forming a ``mist" within the hot halos.  In order to understand the morphology and physical properties of the CGM it is therefore essential to directly map the cold gas emission. \par
In recent years, the ubiquitous detection of bright and extended Ly$\alpha$ emission around quasars at z$>2$ from 
narrow-band imaging and integral field spectroscopy has opened up the possibility to directly study the morphology, kinematics and density distribution of the CGM and IGM at high-redshift on both small (kpc) and large scales (Mpc) \citep[e.g.][]{cantalupo2014cosmic,hennawi2015quasar, borisova2016ubiquitous,fumagalli2017witnessing,arrigoni2019qso, cai2019evolution,martin2019multi,o2019flashes,fossati2021muse}.
In addition to the availability of new instruments such as MUSE \citep[see][]{bacon2010muse} and KCWI \citep[see][]{morrissey2018keck}, the success of these surveys targeting the Ly$\alpha$ emission is due to the intrinsic luminosity of the line expected when dense gas is ionized directly by a quasar. 

However, directly translating the observed Ly$\alpha$ emission into gas densities and kinematics is challenging without knowing the detailed contribution of the individual Ly$\alpha$ emission mechanisms to the total and because of the resonant nature of the Ly$\alpha$ transition. 
In addition to being produced by the recombination of a free electron with ionized hydrogen, Ly$\alpha$ photons can also be produced by collisional excitation between a free electron and a neutral hydrogen atom or by ``continuum pumping". In the latter process, a photon produced by, e.g. the accretion disk of the quasar, and doppler-shifted to the frequency of Ly$\alpha$ radiation in the rest frame of a hydrogen atom in the CGM or IGM would be re-absorbed and re-emitted (``scattered") by the atom, effectively contributing to the extended Ly$\alpha$ emission around the quasar. Both collisional excitation (CE) and ``continuum pumping" (CP) require the presence of neutral hydrogen atoms in the CGM/IGM and very favourable conditions in terms of free electron temperatures or gas kinematics \citep[e.g.][]{fardal2001cooling, cantalupo2008mapping,  cantalupo2014cosmic, dijkstra2019physics, pezzulli2019high}. From the Ly$\alpha$ line alone, it is however difficult to estimate the possible contribution of these two emission mechanisms compared to recombination radiation. 
For the typical densities of the CGM/IGM up to interstellar medium densities, the H$\alpha$ emission of hydrogen or hydrogen-like atom, such as \ion{He}{II}, is expected to be effectively produced by atomic recombinations only, due to the non-resonant nature of this line. Note also that the collisional excitation rates to the 3s, 3p and 3d levels of \ion{H}{I} (which result in H$\alpha$ photons or Ly$\beta$ photons) are about 20 times smaller than the collisional excitation rate to the 2p level (which result in Ly$\alpha$ photons) at $T\sim10^4$ K \citep[see e.g.,][]{osterbrock1962escape, spitzer1978review,giovanardi1987numerical, gould1996imaging, cantalupo2008mapping} which would result in Ly$\alpha$/H$\alpha$ flux ratios $> 100$.  
The detection and measurement of hydrogen H$\alpha$ or \ion{He}{II} 1640 flux can therefore be used to constrain the emission mechanism and therefore gas densities. Unfortunately, hydrogen H$\alpha$ emission is undetectable from the ground at z$>$3, where the vast majority of quasar Ly$\alpha$ nebulae have been detected so far, because of atmospheric absorption and emission lines and requires space-based facilities such as JWST. \ion{He}{II} 1640 emission, which falls in the optical for z$>$3, has proven difficult to detect \citep[e.g.][]{langen2019neb, borisova2016ubiquitous, fossati2021muse} with typical \ion{He}{II} $\lambda$1640/Ly$\alpha$ $<$ 0.1 (instead of 0.3 expected from fully ionized helium and hydrogen and recombination radiation). This small \ion{He}{II}/Ly$\alpha$ ratio is expected, even in the pure recombination scenario, if CGM/IGM gas has a broad density distribution with tails extending to values at which helium becomes completely self-shielded to the \ion{He}{II}-ionizing photons ($\sim$54eV) produced by the quasar, as demonstrated by \citep{cantalupo2019large}. The detection of hydrogen H$\alpha$ emission becomes therefore the main requirement for a secure confirmation of the recombination origin of the Ly$\alpha$ emission. 

The Slug Nebula at z$\sim2.3$ \citep{cantalupo2014cosmic} is so far the only quasar Ly$\alpha$ nebula, to our knowledge, for which H$\alpha$ emission has been searched for and detected \citep[e.g.][]{leibler2018detection}. In particular, using MOSFIRE slit-spectroscopy, H$\alpha$ was detected coincident with the brightest Ly$\alpha$ emission both spatially (within the MOSFIRE slit) and kinematically. Unfortunately, the presence of a bright sky line hampered the measurement of the spatial and spectral profile of the emission, limiting our ability to constrain both the morphology of the emission on kpc scale and its kinematic properties. Nonetheless, by heavy smoothing both spatially and spectrally a high signal-to-noise measurement has been possible showing a Ly$\alpha$/H$\alpha$ flux ratio of 5.5$\pm$1.1. As discussed by \cite[][]{leibler2018detection}, such a line ratio is fully compatible with a recombination origin of Ly$\alpha$ emission (including slit-loss effects due to local Ly$\alpha$ radiative transfer) and strongly excludes any significant contribution from ``photon-pumping/scattering" or "collisional excitation/cooling".

Can this conclusion be applied generally to Ly$\alpha$ nebulae or is it only the case for the Slug Nebula? In order to address this question, it is essential to build up a larger sample of nebulae with observations of H$\alpha$ emission. Moreover, in order to take full advantage of the non-resonant nature of H$\alpha$ emission as a kinematic tracer, it is essential to avoid overlapping with sky lines. Here we present results of a pilot project, using both KCWI and MOSFIRE specifically designed to meet the above requirements. The successful results of this project, as described in detail below, lay the foundation for building up a larger sample of extended H$\alpha$ emission around quasars in future surveys.  

The paper is structured as follows. In Section~\ref{sec:lyaobservations} we describe our survey strategy and target selection. In Section~\ref{sec:Lyaobs} we present the KCWI observations for extended Ly$\alpha$ emission. %within a optimal redshift range for subsequent MOSFIRE follow-up searches for H$\alpha$ emission.
The MOSFIRE observations are described in Section~\ref{sec:Haobs}. In Section~\ref{sec:comp}, we present a detailed analysis of Ly$\alpha$ and H$\alpha$ emission. Finally, we discuss the nebular emission mechanisms in Section~\ref{sec:discussion}. Conclusions and future prospects are reported in Section~\ref{sec:conclusion}.

\section{Survey strategy and Target selection}%Lyman-$\alpha$ Observations} 
\label{sec:lyaobservations}

The goal of our survey is the measurement of large-scale H$\alpha$ emission around quasars at high spatial and spectral resolution. For this reason, it is essential that the expected H$\alpha$ emission falls in a wavelength range with high atmospheric transparency in the IR without the presence of any significant sky line. Based on the expected IR sky emission at Mauna Kea, we estimated that the optimal emission redshift range that satisfies the requirement above would be $2.25<z<2.27$.

Given the need to place the slit for MOSFIRE H$\alpha$ observation, a prior knowledge is required of the morphology and extension of the possibly emitting gas, as traced, e.g. by Ly$\alpha$ emission. Moreover, in order to maximize detectability and to address the main scientific question of our project, the Ly$\alpha$ emission should be sufficiently bright that even an H$\alpha$ non-detection would place useful constraints on the emission mechanism.  

Unfortunately, no previously discovered quasar Ly$\alpha$ nebula with bright and extended emission falls in the optimal redshift range described above (within the window of observability for our survey). For this reason, and taking advantage of the demonstrated ubiquity of Ly$\alpha$ nebulae around quasars, we selected new fields around quasars with redshifts within the selected range as targets for KCWI Ly$\alpha$ emission observations. In the expectations that the brightest quasars would have the most luminous nebulae, although so far only a weak trend with luminosity has been found \citep[see][]{mackenzie2021faintqso}, we restricted our selection to quasars with magnitudes i $< 18.5$.
Finally, we made a selection in Right Ascension (RA) to guarantee that the targets would be visible during scheduled runs with KCWI (2019 September 26, 27, 28) and MOSFIRE (2019 November 4). 

Based on these criteria, we selected 3 SDSS quasars from the 12$^{th}$ data release \citep{paris2017sloan} for the KCWI observations as described in Table~\ref{tab:sample}. They are listed from left to right in order of preference regarding the selection criteria. 

The catalogue redshifts are $z=$2.2594 $\pm$ 0.002, 2.2643 $\pm$ 0.001 and 2.26329 $\pm$ 0.0001 for J2316, J0030 and J0010 respectively, which were estimated using the method explained by the \href{https://www.sdss.org/dr12/algorithms/redshifts/}{SDSS redshift and classification documentation}\footnote{\href{https://www.sdss.org/dr12/algorithms/redshifts/}{https://www.sdss.org/dr12/algorithms/redshifts/}}.

\begin{table}
\caption [Sample data] {Selected targets for our observing runs. The targets were observed during 3 nights, one for each night. All nights had similar sky conditions with a mean seeing between 0.54" and 0.94" and no clouds. }
\vskip 1mm

\resizebox{.5\textwidth}{!}{

\begin{threeparttable}
\begin{tabular}{l| c c c}

\hline \\ [0.01ex]
                                        &       J2316+09 &     J0030+05      &      J0010+06       \\ [1ex]
\hline \\ [0.01ex]
Right Ascension          & 23:16:49.5 &   00:30:21.8     & 00:10:5.7    \\ [0.3ex]
Declination              & +09:06:34.82  &    +05:30:52.93     & +06:17:20.63    \\ [0.3ex]
$z_{\tiny \textrm{SDSS}}$ \tnote{1} & 2.2633 $\pm$ 0.0001  &   2.2644 $\pm$ 0.0001      & 2.2595 $\pm$ 0.0002   \\ [0.3ex]
Magnitude \textit{i}-band &   18.01         &   18.14           &    18.49                  \\ [0.3ex]
Exp. Time KCWI\tnote{2}     &  40 min       &   40 min                 &  40 min       \\ [0.3ex]
Mean seeing     & 0.54" & 0.49" & 0.94"  \\ [0.3ex]
%Class \tnote{3}          & BL     &   BL    & BL            \\ [1ex]
\hline
\end{tabular}
\vskip 4mm
\begin{tablenotes}
    \item[1] SDSS redshift obtained from rest-frame principal-component analyses (PCA), see \cite{bolton2012spectral} and \href{https://www.sdss.org/dr12/algorithms/redshifts/}{SDSS redshift and classification documentation}.
    %\footnote{\href{https://www.sdss.org/dr12/algorithms/redshifts/}{https://www.sdss.org/dr12/algorithms/redshifts/}}
    \item[2] The exposure time of 40 min total is composed of 4 individual exposures of 10 min each.  
\end{tablenotes}
\end{threeparttable}
} 
\label{tab:sample}
\end{table}

\section{Lyman-\texorpdfstring{$\alpha$}{alpha} Observations}
\label{sec:Lyaobs}

In this Section, we present our Ly$\alpha$ observations using the Keck Cosmic Web Imager (KCWI) integral field spectrograph, including a brief description of the chosen KCWI instrument configuration, data reduction and analysis.  

\subsection{KCWI configuration and observations}
\label{subsec:kcwi}

KCWI is an Integral Field Spectrograph (IFS) installed on the Keck II telescope on Mauna Kea, Hawaii, using a configuration that provides a spectral range $3500-4600\,\rm{\AA}$, which is ideal to search for Ly$\alpha$ emission in our selected redshift range. The design provides configurable spectral resolution from $\rm{R}=1000-20000$ with a Field of View (FoV) up to $20"\times33"$ \citep[for further details see][]{morrissey2018keck}. 
%\par
For our observations, we choose a spectral resolution of $\rm{R}=4100$, corresponding to a FWHM $\sim$ 73 km s$^{-1}$ with the ``medium slicer" which provides a FoV of 16.5" $\times$ 20" and the ``BM'' grating, corresponding to $\sim$132$\times$166 physical kpc$^2$ . The selected configuration is ideal for the purposes of our study providing both a high sensitivity for diffuse and faint emission over a large area and the ability to resolve spectrally narrow line emission. 

The targets were observed over the course of three nights, one during each night. The nights had similar weather conditions and a mean seeing of 0.54", 0.49" and 0.94" for night one, two and three respectively.
The total exposure time on source for each field is 40 min, divided into 4 individual exposures of 10 min each, reaching a depth in SB of about 10$^{-19}$ erg s$^{-1}$cm$^{-2}$arcsec$^{-2}$. The field of view has been offset by 3 arcsec and rotated by 90$^{\circ}$ between individual exposures in order to improve the spatial sampling and to reduce possible systematics due to flat fielding and sky subtraction, as described in more detail in Section \ref{subsubsec:kcwi_datareduction}.

\subsection{Data Reduction}
\label{subsubsec:kcwi_datareduction}

Data reduction has been performed using the Standard KCWI Data Reduction Pipeline (DRP)\footnote{\href{https://github.com/kcwidev/kderp/}{ https://github.com/kcwidev/kderp/}}, released in March 17, 2018 \citep[see][]{morrissey2018keck}. 
Twilight flats taken at the beginning or the ending of the same nights have been used for flat-fielding while wavelength calibration has been performed using a Thorium-Argon lamp. 
For flux calibration, we used the photometric standard star (G191b2b), which was observed during the same nights as the science exposures.
The standard DRP was complemented by custom-made scripts to perform the empirical low-order background subtraction, in which a sky model was obtained from a median-filtered cube. Moreover, all cubes were reduced to vacuum wavelength and corrected for the heliocentric velocity. 
 
Prior to the stacking process, the astrometric solution was corrected by cross-correlating white-light images of the fields (obtained by integrating the cubes along the wavelength direction) with r-band SDSS images.
Pipeline reduced cubes for each exposure were resampled into a common grid with $0.2" \times 0.2" \times 0.5 \AA$ pixel size, using an ``Drizzle'' algorithm implemented with \href{http://montage.ipac.caltech.edu/}{Montage}.\footnote{\href{http://montage.ipac.caltech.edu/}{http://montage.ipac.caltech.edu/}} Details on the above-mentioned custom procedures in addition to the DRP can be found in \cite{chen2021kbss}.
Finally, in order to improve sky subtractions we have used the white-light images of each individual cube to correct for flat-fielding and illumination residuals using a procedure similar to the tool \textsf{CubeFix}, part of the \textsf{CubExtractor} package \citep{cantalupo2019large}.

\subsection{Extraction of the Nebular Emission}
\label{subsec:extraction}

\subsubsection{Quasar PSF and source continuum removal}
\label{subsubsec:psfsub3D}

As commonly done in previous studies, the extraction of faint nebular emission around quasars requires the prior removal of the quasar Point Spread Function (PSF) which would otherwise dominate over the diffuse emission, especially in the central regions.
For this task, we used the \textsf{CubePSFSub} procedure (part of the \textsf{CubExtractor} package) with the following parameters: a spectral width of 100 layers for the pseudo-broad-band images, a spectral PSF size of 500 layers and a spatial PSF rescaling-region of two pixels, corresponding to 0.4". Note that we mask the range of layers where extended emission is expected to avoid affecting the empirically reconstructed PSF by the nebular emission and hence to avoid oversubtraction. 
These parameters are different than the typical choices for MUSE datacubes \citep[e.g.][]{borisova2016ubiquitous, cantalupo2014cosmic, cantalupo2019large, arrigoni2019qso} in order to take into account the different spectral and spatial resolution of KCWI and other differences between KCWI and MUSE. Note that a central region with size 0.4" times 0.4" is assumed to be dominated by the quasar PSF and it is used for the empirical rescaling procedure of \textsf{CubePSFSub} \citep[see][for details]{cantalupo2019large}. This area was masked in our scientific analysis as commonly done in previous studies. 

After PSF subtraction, we removed the continuum from other sources in the field using the \textsf{CubExtractor} procedure \textsf{CubeBKGSub} performing median filtering with the following parameters: a bin size of 200 spectral pixel (corresponding to 100\AA) and a smoothing radius of two spectral pixel. As mentioned for the PSF subtraction, the spectral range of extended emission has been masked to avoid oversubtraction.

\subsubsection{Detection and extraction of extended Ly$\alpha$ emission} 
\label{subsubsec:optextr}

The detection and extraction of the nebular emission was performed using the \textsf{CubExtractor} software \citep[see][for a description]{cantalupo2019large}. 
Prior to detection,  a 2D Gaussian smoothing kernel with radius=1 pixel was applied to the data and corresponding variance cube. Then, objects from the 3D cubes were extracted if they contain at least 200 connected voxels (3D pixels) with signal to noise (S/N) threshold $\geq$ 2.2 per voxel. 
The information about the voxels associated with the detected sources is stored in a 3D segmentation mask that is then used to obtain optimally extracted images similarly to previous studies, i.e. by integrating along the wavelength direction over only the voxels associated with the sources \citep[e.g.][]{borisova2016ubiquitous, cantalupo2014cosmic, cantalupo2019large, arrigoni2019qso}. In order to improve the final S/N,  a further spatial boxcar filter smoothing with a size of 0.4" $\times$ 0.4" has been applied to the cubes before integrating along the wavelength direction.  

\begin{figure*}
\centering
  \begin{subfigure}{0.33\textwidth}
    \centering  
    \includegraphics[width=\textwidth]{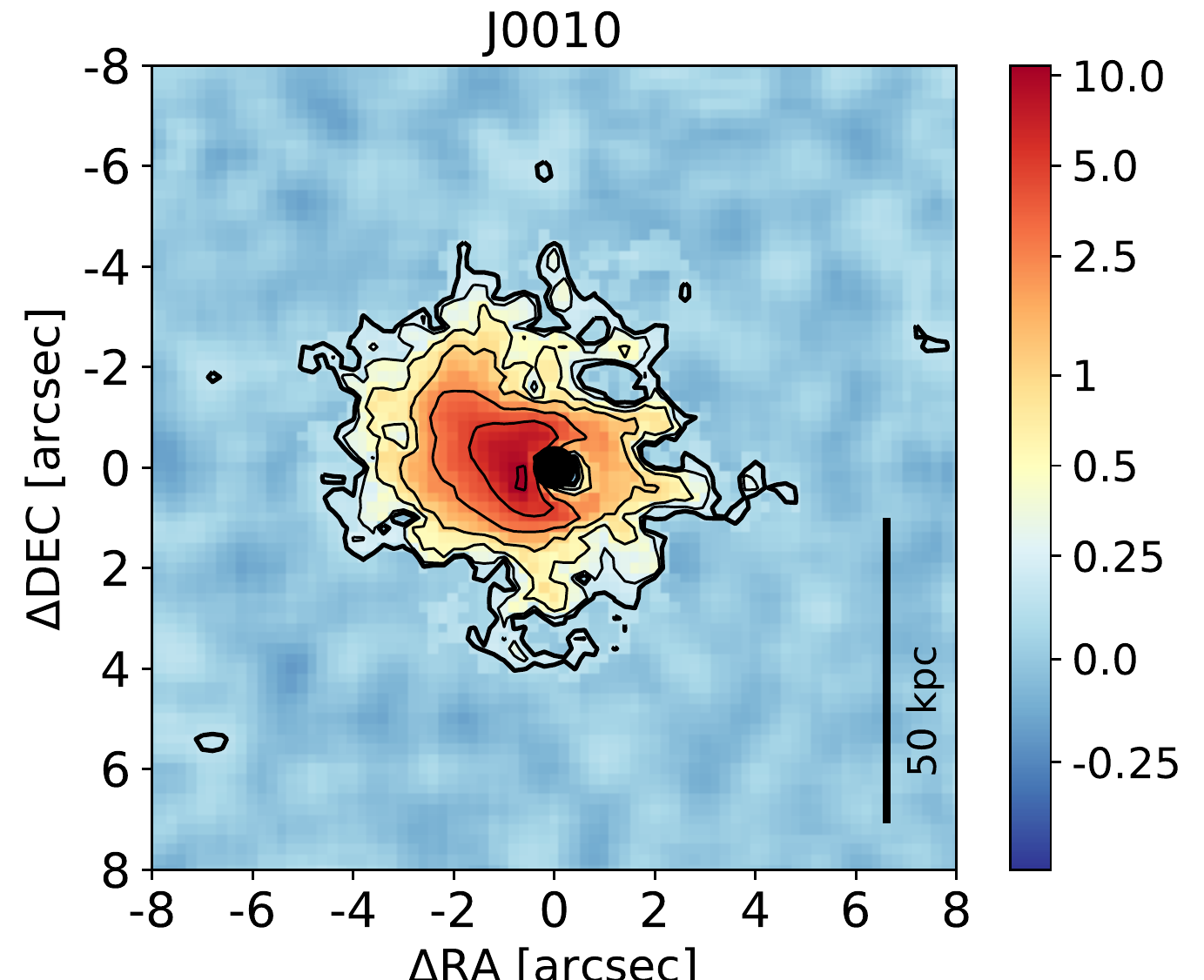}
  \end{subfigure}
 \hfill
  \begin{subfigure}{0.33\textwidth}
    \centering
    \includegraphics[width=\textwidth]{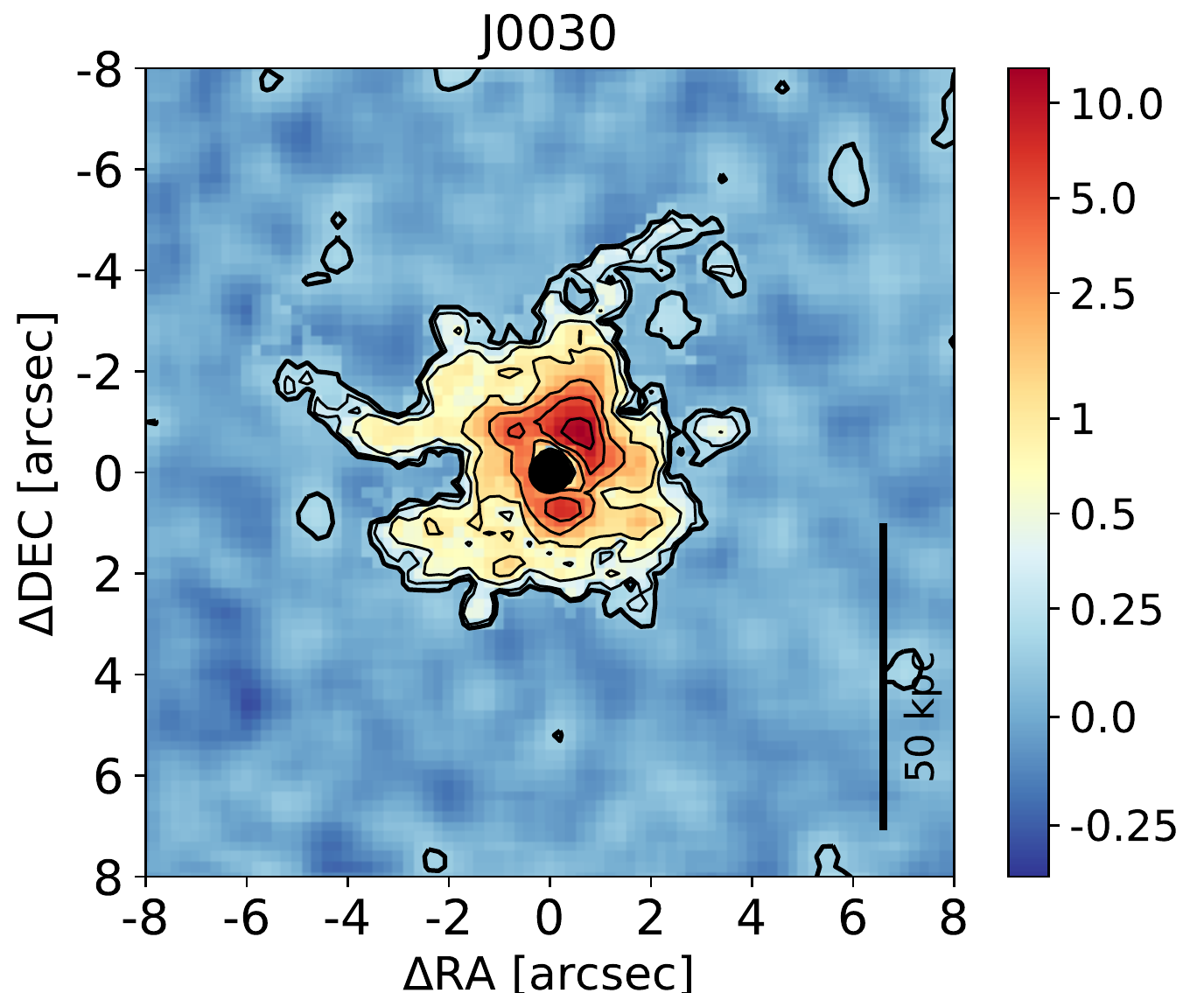}
  \end{subfigure}
 \hfill
 \begin{subfigure}{0.33\textwidth}
    \centering
    \includegraphics[width=\linewidth]{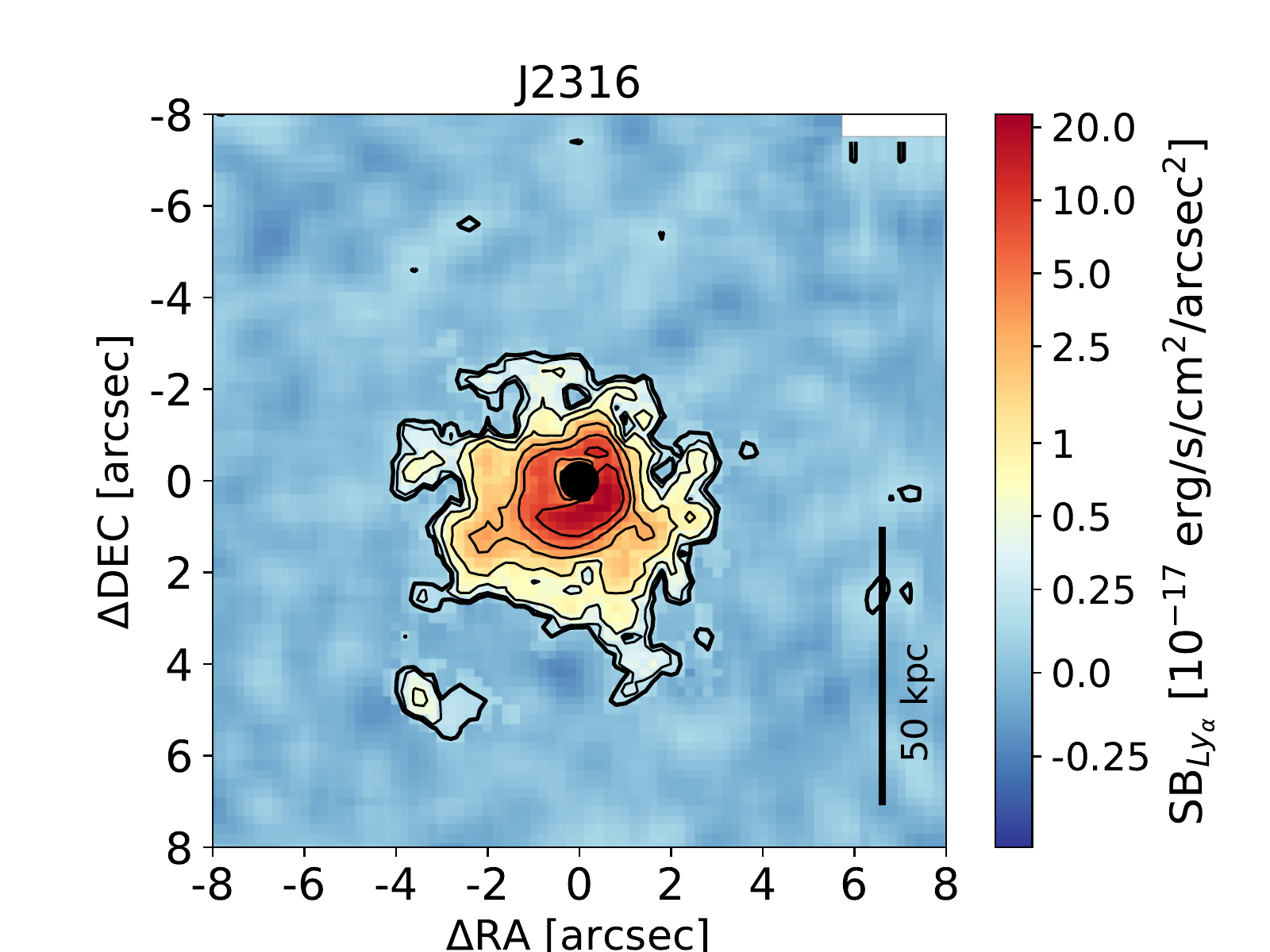}
 \end{subfigure}

\caption[Surface brightness map of the Ly$\alpha$ nebulae]{Surface brightness (SB) Ly$\alpha$ emission maps. The nebular emission has been optimally extracted using the 3D segmentation mask obtained by \textsf{CubEx} with a SNR threshold of 2.2, (see Section~\ref{subsubsec:optextr} for more details), and was smoothed using a box car filter with a width of $\sim$0.4" %[THIS BOXCAR FILTER SIZE IS NOT CONSISTENT WITH WHAT IS WRITTEN IN THE TEXT]. 
Note that the color scale is logarithmic. 
A noise layer at the position of the flux weighted centroid of the nebula is shown in the background for visualization purposes.
The zero position in the x and y dimensions correspond to the quasar position used in the PSF subtraction. The black filled circle masks the area used for the PSF rescaling (see Section \ref{subsubsec:psfsub3D}) which is thus unusable for our scientific analysis. 
The thick outermost contour corresponds to a SB level of 0.15$\times$10$^{-17}\mathrm{erg} \ \mathrm{s}^{-1} \mathrm{cm}^{-2} \mathrm{arcsec}^{-2}$. The subsequent contours represent levels of 0.25, 0.5 1, 2.5, 5., 10. in units of 10$^{-17}\mathrm{erg} \ \mathrm{s}^{-1} \mathrm{cm}^{-2} \mathrm{arcsec}^{-2}$ and are also presented by the colorbar tick marks. 
 
}
\label{fig:SBmapLya}
\end{figure*}

Following this procedure, we detected bright Ly$\alpha$ nebulae around all 3 targets at a similar redshift with respect to the quasars. In 2 cases the nebulae reached a projected size up to 50 kpc, corresponding to about 6 arcsec in the sky for the given redshift. 
We present the optimally extracted images of the detected Ly$\alpha$ nebulae within each field in Figure~\ref{fig:SBmapLya} and report the key physical quantities in Table~\ref{tab:sample_properties}. Continuum images of these fields (before and after the quasar PSF subtraction) are presented in the Appendix~\ref{app:WLimages}. 
The Ly$\alpha$ emission detected around the three Quasi-Stellar Objects (QSOs) has integrated fluxes between F$_{ \textrm{\tiny Ly}\alpha}$= (40.6 $\pm$ 0.3) $\times$ 10$^{-17}\mathrm{erg} \ \mathrm{s}^{-1} \mathrm{cm}^{-2}$ and F$_{ \textrm{\tiny Ly}\alpha}$= (66.9 $\pm$ 0.3) $\times$ 10$^{-17}\mathrm{erg} \ \mathrm{s}^{-1} \mathrm{cm}^{-2}$ and measure average surface brightness (SB) between SB$_{\tiny \textrm{Ly}\alpha}$= (0.8 $\pm$ 0.3) $\times$ 10$^{-17}\mathrm{erg} \ \mathrm{s}^{-1} \mathrm{cm}^{-2} \mathrm{arcsec}^{-2}$ and SB$_{\tiny \textrm{Ly}\alpha}$= (1.6 $\pm$ 0.3) $\times$ 10$^{-17}\mathrm{erg} \ \mathrm{s}^{-1} \mathrm{cm}^{-2} \mathrm{arcsec}^{-2}$, as summarized in Table~\ref{tab:sample_properties}.
The fluxes were calculated by integrating the surface brightness over the area defined by the 3D segmentation mask obtained by CubExtractor. 
The peak SB is typically reached close to the QSO position. 
The parameter $z_{\tiny \textrm{SDSS}}$ refers to the quasar redshift listed in the SDSS catalog, whereas $z_{\tiny \textrm{nebula}}$ indicates the nebular redshift calculated using the peak in the integrated Ly$\alpha$ spectrum. 
%EXPLAIN WHY Ha was NOT used for peak estimation
As observed in previous studies, these two measurements differ from each other. In particular, we observe a positive offset of $\sim$510 km s$^{-1}$ and $\sim$600 km s$^{-1}$ for J0010 and J0030 respectively, between the Ly$\alpha$ redshift and the quasar SDSS redshift, whereas J2316+09 shows instead a blueshift of a $\sim$570 km s$^{-1}$. \par
Consistent with previous observations of Ly$\alpha$ nebulae at similar redshift \citep[e.g.][]{cai2017discovery, o2019flashes}, our detected nebulae present very asymmetric morphology. Such asymmetry is especially prominent for J0010+06 for which the region with the highest SB is elongated towards north-east (NE).

J0010+06 contains the most extended and luminous Ly$\alpha$ filament at a redshift that would place the associated H$\alpha$ emission in the best part of the infrared spectrum in terms of absence of sky lines. The extended emission is not associated with any visible continuum object (see Figure~\ref{app:WL} in the Appendix~\ref{app:WLimages}) ensuring that the detected emission is indeed circum-galactic or intergalactic. 

For these reasons, J0010+06 was selected as the target for a MOSFIRE follow-up search for extended H$\alpha$ emission as described below.

\begin{table*}
  \centering

  \caption[Main physical properties of the observed fields]{Main properties of the giant Ly$\alpha$ nebulae detected with KCWI around the quasar fields J2316+09, J0030+05, and J0010+06. }
\label{tab:J0010facts}

\begin{threeparttable}
\begin{tabular*}{0.99\textwidth}{l@{\extracolsep{\fill}} l ccc }
\hline
\hline \\ [0.01ex]
     &                                                                       &   J0010+06  &   J0030+05  &  J2316+09           \\ [1ex]
\hline \\ [0.01ex]

Flux & [10$^{-17}\mathrm{erg} \ \mathrm{s}^{-1} \mathrm{cm}^{-2}$]   & (48.1 $\pm$ 0.3) &  (40.6 $\pm$ 0.3) & (66.9 $\pm$ 0.3)  \\ [0.3ex]
Projected Area \tnote{a} & [arcsec$^2$]                                           & $\sim$ 46     &  $\sim$ 48    &  $\sim$ 43       \\ [0.3ex]
Average SB & [cgs] \tnote{b}                                         & (1.0 $\pm$ 0.2) & (0.8 $\pm$ 0.3) & (1.6 $\pm$ 0.3)  \\ [0.3ex]
Peak SB & [cgs]                                                      & (10.8 $\pm$ 0.2) & (12.9 $\pm$ 0.3) & (22.6 $\pm$ 0.2)   \\ [0.3ex]
$z_{\tiny \textrm{SDSS}}$ &                                          & 2.2633 $\pm$ 0.0001 & 2.2644 $\pm$ 0.0001 & 2.2595 $\pm$ 0.0002 \\ [0.3ex]
$z_{\tiny \textrm{nebula}}$ \tnote{c} &                              & 2.257 $\pm$ 0.005 & 2.27 $\pm$ 0.01 & 2.26 $\pm$ 0.01      \\ [0.3ex]
\hline

\end{tabular*}

\vskip 4mm
\begin{tablenotes}
    \item[a] The spatial extent refers to the area bound by the outer contour of the binary mask obtained with \textsf{CubEx} for which a SNR threshold of 2.2 was used.  
    \item[b] cgs: 10$^{-17}\mathrm{erg} \ \mathrm{s}^{-1} \mathrm{cm}^{-2} \mathrm{arcsec}^{-2}$
    \item[c] The redshift was calculated via the equation $z = \rm{\lambda_{observed}/\lambda_{vacuum}} - 1$ where $\rm{\lambda_{observed}}$ is the position of the maximum of the integrated Ly$\alpha$ flux of the nebular emission in vacuum and $\lambda_{\textrm{vacuum}}$ is the vacuum wavelength. The uncertainty has been estimated by the FWHM of the integrated Ly$\alpha$ flux of the nebula emission. 
\end{tablenotes} 
\end{threeparttable}
\label{tab:sample_properties}
\end{table*}

%%%%%%%%%%%%FOV
% PART    2%                      
%%%%%%%%%%%%%%%%%%%%%
% H-A OBSERVATIONS %
%%%%%%%%%%%%%%%%%%%%%
% going from MOSFIRE facts to its results 
\section{Hydrogen H-\texorpdfstring{$\alpha$}{alpha} Observations}
\label{sec:Haobs}

The following sections describe our MOSFIRE observation aimed at the detection of extended H$\alpha$ emission near J0010+06. 
In particular, we describe here the instrument configuration, the data reduction, analysis and final results.

\subsection{MOSFIRE configuration}
\label{subsec:followup}
The Multi-Object Spectrograph for Infrared Exploration (MOSFIRE) is a slit spectrograph which is particularly suited for the detection of faint emission in the infrared and installed on the Keck I telescope on Mauna Kea, Hawaii (described in detail in \cite{mclean2012mosfire}; \cite{steidel2014strong}). 
The spectrograph covers a wavelength range of 9700 to 24100 $\rm{\AA}$, a Field of View of 6.1' $\times$ 6.1' and has a resolving power of $\rm{R}=3660$ for a slit width of 0.7" (0.508 mm). 
Order-sorting filters provide spectra that covers the K, H, J or the Y band by selecting the 3$^{\small \textrm{rd}}$, 4$^{\small \textrm{th}}$, 5$^{\small \textrm{th}}$ or 6$^{\small \textrm{th}}$ order respectively.  
The K-band supports a range at center of 19450 - 23970$\AA$, which can be extended down to 19210$\AA$ or up to 24060$\AA$ by optimizing the slit mask design; the dispersion is ~2.1691$\AA$/pixel. 
Given the redshift range targeted in our study, 
%and a vacuum wavelength of $\lambda_{ \textrm{vac}}$=6564.61$\AA$ for the H$\alpha$ line, 
we expect to detect H$\alpha$ near $\lambda_{\textrm{obs}}$ = (1+$z$)$\lambda_{\textrm{vac}}\sim$ 21450$\AA$, where $\lambda_{ \textrm{vac}}$ = 6564.61$\AA$, and hence selected the K-band configuration. 
Taking advantage of the configurable MOSFIRE slit bars, we chose a single slit with a width of 1" and a length of 2', centered on the position of the J0010+06 quasar. 
With a 1" slit, the spectral resolution power is R$\sim$2600 ($\Delta$v $\sim$ 115 km s$^{-1}$). This resolution maximizes the detectability of the expected diffuse, and spectrally narrow H$\alpha$ emission allowing us to detect possible multiple velocity components \cite{cantalupo2019large}, \cite{price2019mosdef}.
The position angle (PA) of the slit (80$^{\circ}$) has been chosen in such a way to cover the brightest part of the J0010+06 nebula and further optimize the H$\alpha$ detectability, as can be verified in Figure \ref{fig:SBmapLya2}. In case of non-detection, targeting the brightest part of the Ly$\alpha$ nebula would give us in any case a Ly$\alpha$/H$\alpha$ lower limit which would be high enough to address our main scientific question. 

The target J0010+06 was observed on the night of 2019 November 3. The sky conditions were good at the time of observations with a mean seeing of 0.65". A total exposure time of 5400s (380x30) was obtained with two separate exposure sequences of 45 min each. During both exposures, we applied a ABAB dither sequence of short individual exposures for which the frame was moved perpendicular to the slit between position A and B every 3 min, as further explained in the next section.

\subsubsection{MOSFIRE Data Reduction}
\label{subsubsec:mosfire_datareduction}

Data reduction, including flat-fielding, wavelength calibration and sky subtraction, was performed using the MOSFIRE data reduction pipeline (DRP), publicly available and developed by the instrument team, see \cite{mclean2012mosfire}\footnote{\href{https://keck-datareductionpipelines.github.io/MosfireDRP/}{MOSFIRE DRP Documentation: https://keck-datareductionpipelines.github.io/MosfireDRP/}}.
Flux calibration as well as astrometric solutions were performed using ``MOSPEC'', an IDL based 2D spectral analysis tool developed specifically for the MOSFIRE instrument, described in detail by \cite{strom2017nebular}. 
The further reduction was performed in 2 stages. 
During the first stage, we subtracted pairwise interleaved, dis-registered stacks, which are obtained using the ABAB dither sequence of short individual exposures and combining frames taken in positions A and B separately. This routine ensures a good background subtraction, obtained by a simple subtraction (i.e., A-B or B-A) minimizing possible sky variations during the night. Residuals are then removed by fitting a 2D b-spline model to the remaining background, with a method similar to \cite{kelson2003optimal}. The DRP further produces flat-fielded, wavelength calibrated, rectified and stacked 2D spectrograms for each slit on a given mask. The 2D wavelength solutions for the K-band were obtained using a combination of the night sky and Neon arc lamp spectra, automatically executed by the DRP.  
All spectra were reduced to vacuum wavelength and corrected to the heliocentric frame. The two individual measurements, which were taken on 2 different nights, were combined using inverse-variance weighting to form the final 2D spectra.
Finally, using MOSPEC, 1D spectra together with their associated 1$\sigma$ error spectra, were extracted from the final background-subtracted frame. Flux calibration was accomplished using AO stars observed on the same nights. 
%More detailed information can be found in \cite{steidel2014strong}.

\subsubsection{Quasar continuum and H$\alpha$ emission subtraction}
\label{subsubsec:psfsub2D}

Similar to the Ly$\alpha$ emission, the detection and study of faint H$\alpha$ intergalactic emission close to the quasar requires the prior removal of the quasar continuum and broad H$\alpha$ emission. 
For this purpose, we developed a quasar line-spread-function (LSF) subtraction tool specifically for the MOSFIRE two dimensional data format. The algorithm follows a similar approach as \textsf{CubPSFSub} (see Section~\ref{subsubsec:psfsub3D} and the procedures used by \cite{o2019flashes}), as we describe in detail below.
For each wavelength position in the 2D spectra, a 1D spatial emission profile is created by summing over 46$\AA$ centered around the current wavelength masking bright sky residuals. This wavelength window was chosen in order to obtain a sufficient signal-to-noise that is sensitive to spectral fluctuations in the QSO continuum and broad emission lines.

The 1D spatial emission profiles at each wavelength were then fitted with a Moffat profile, rescaled in order to match the flux contained within a radius of 0.5" \footnote{Note that the rescaling region for the MOSFIRE data is almost twice as large as the one for the KCWI data. This difference is due to higher noise fluctuations in the MOSFIRE data such that selecting a median out of 5 pixel as rescaling factor leads a better reduction that averaging over 2 pixel as in the case for KCWI. } around the quasar and then subtracted from the data.
After a first passage, we iteratively re-apply our procedure by masking any detectable line emission features. 

%%bigger?
\begin{figure}
  \centering
  \includegraphics[width=.9\linewidth]{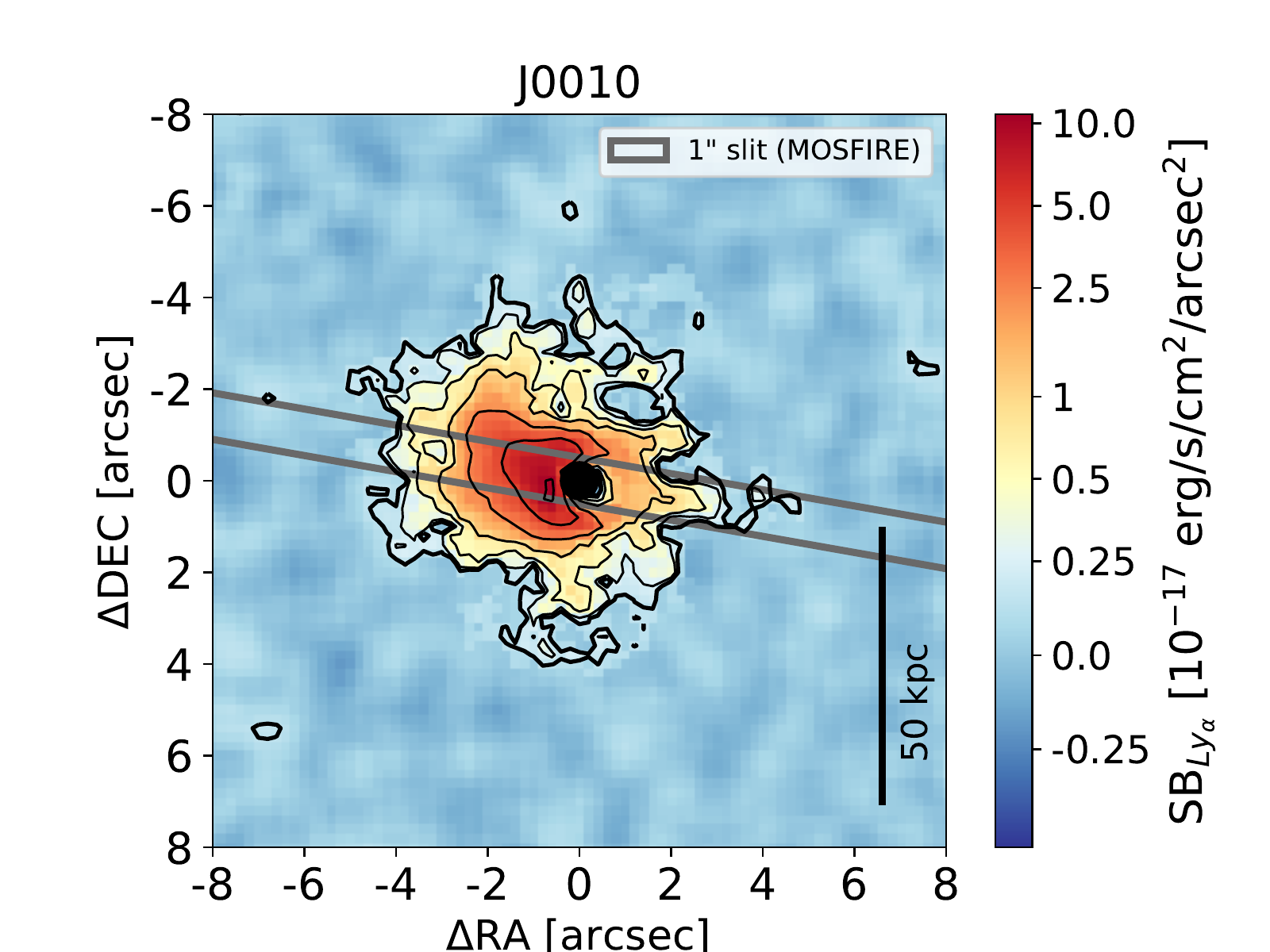}
  \caption{The MOSFIRE slit position for J0010 (gray lines). The background Ly$\alpha$ map is the same as in Figure \ref{fig:SBmapLya}. The slit has been placed at 80$^{\circ}$ from the vertical towards the NE direction such that it covers the brightest part of the Ly$\alpha$ emission and has a width of 1". The positive direction along the slit is defined as NE.}
  \label{fig:SBmapLya2}
\end{figure}

\subsection{The H-\texorpdfstring{$\alpha$}{alpha} Emission of the Nebula}
\label{subsec:Haresults}

\begin{figure}
\includegraphics[width=.95\linewidth]{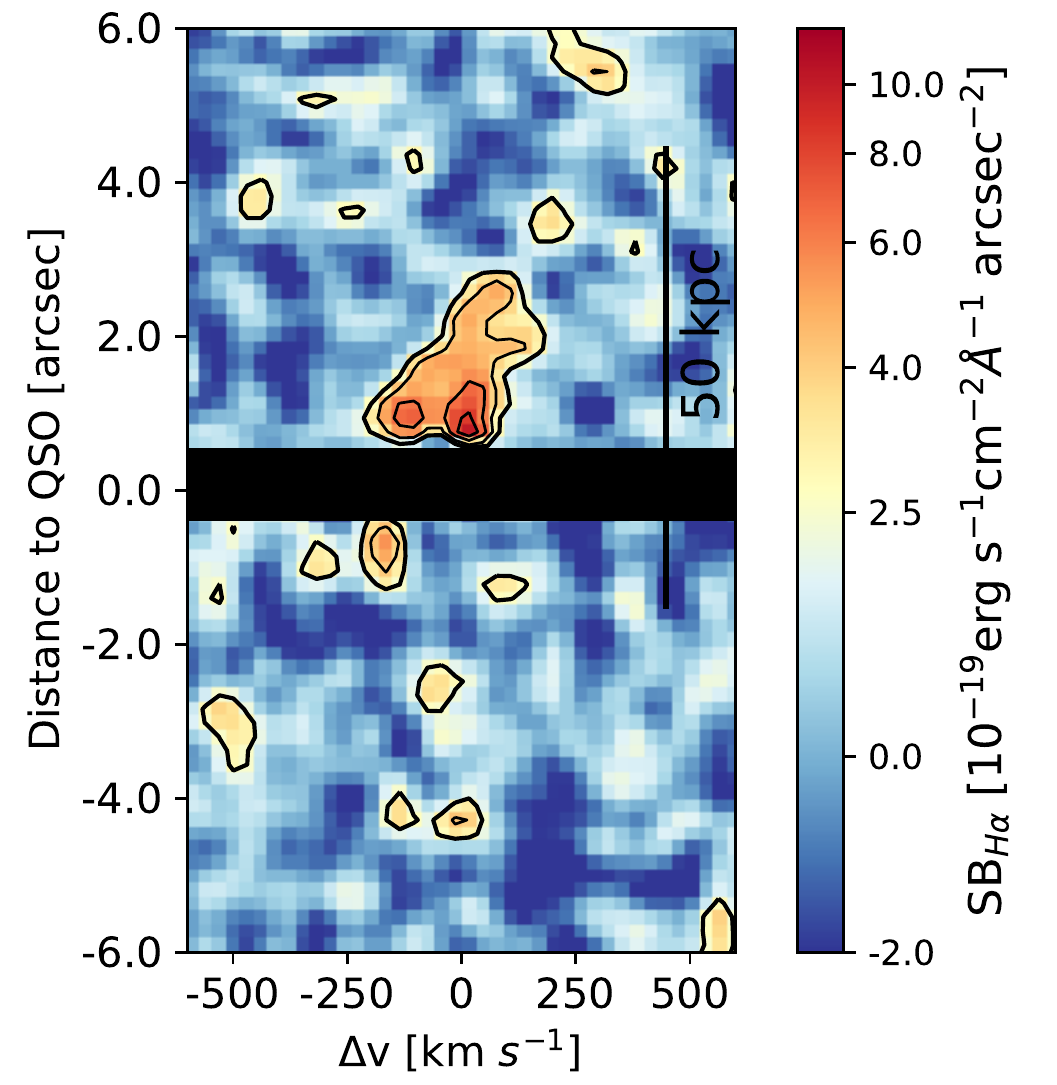}
\caption[H$\alpha$ flux density along the slit]{The flux density of H$\alpha$ along the slit, where the y-axis indicates the spatial distance to the quasar center and the zero represents the center calculated as described in Section~\ref{subsubsec:psfsub2D}. The black bar covers the rescaling region used for the PSF subtraction and can hence not be used for scientific analysis. The x-axis indicates the line-of-sight velocity relative to the redshift of the Ly$\alpha$ nebula in Figure \ref{fig:SBmapLya}. Hereafter, the zero velocity is defined as the peak Ly$\alpha$ emission in the integrated spectra. The contour lines follow constant SB levels of the emission of 2.5, 4.0, 6.0, 8.0 and 10.0 $\times$ 10$^{-19}$ erg s$^{-1}$cm$^{-2}\AA^{-1}$arcsec$^{-2}$. }
\label{fig:2DmapHa}
\end{figure}

Figure~\ref{fig:2DmapHa} shows the MOSFIRE slit 2D spectrum on J0010+06 around the expected H$\alpha$ wavelength position (converted to velocity) and after the subtraction of the quasar emission. The zero velocity corresponds to the redshift of the Ly$\alpha$ peak of the nebula. The central regions, used for the rescaling and subtraction of quasar emission are masked and not used for our scientific analysis. 
H$\alpha$ emission at high signal-to-noise level (above a SNR=3 per spectral and spatial resolution element) is detected near zero velocity and extending by $\sim$2.5", corresponding to about 20 kpc, from the quasar (see also Figure~\ref{app:A1} for a clearer indication of the extent) along the NE side of the slit (denoted by positive y-axis values). This emission is not associated with any visible continuum source or galactic companions to the quasar (Appendix \ref{app:WLimages}). 
Emission is also present on the opposite side at about 1.5" from the quasar. 

We point out, however, that this feature is accompanied by a neighboring negative region of similar size and magnitude (covered under the black rescaling region mask in Figure \ref{fig:2DmapHa}), thus we cannot exclude the possibility that it is an artifact of the PSF subtraction.  %\par

\begin{figure*}
\centering
\begin{subfigure}{0.49\textwidth}
    \centering  
    \includegraphics[width=\textwidth]{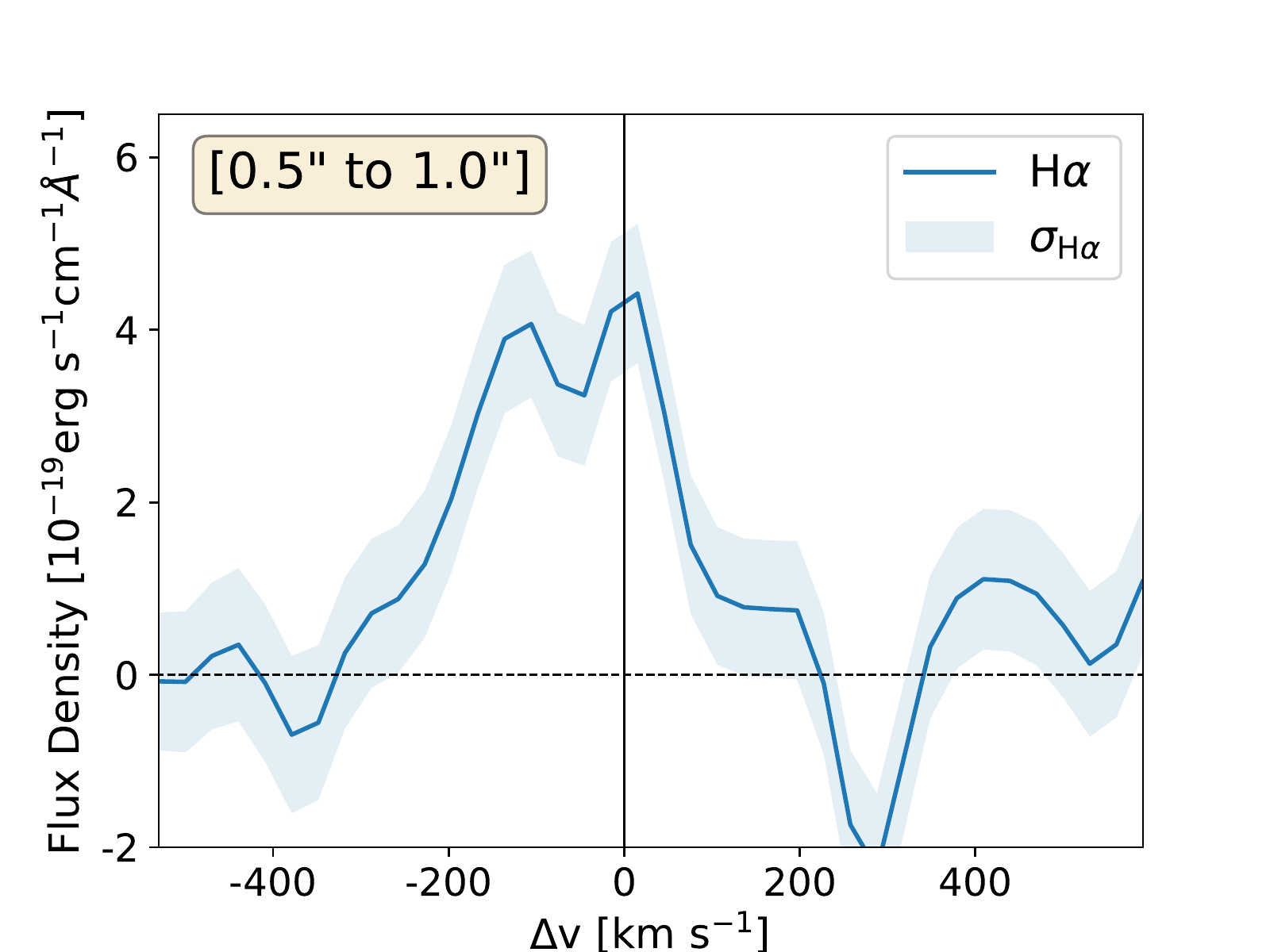}
\end{subfigure}
  \hfill
\begin{subfigure}{0.49\textwidth}
  \centering
  \includegraphics[width=\textwidth]{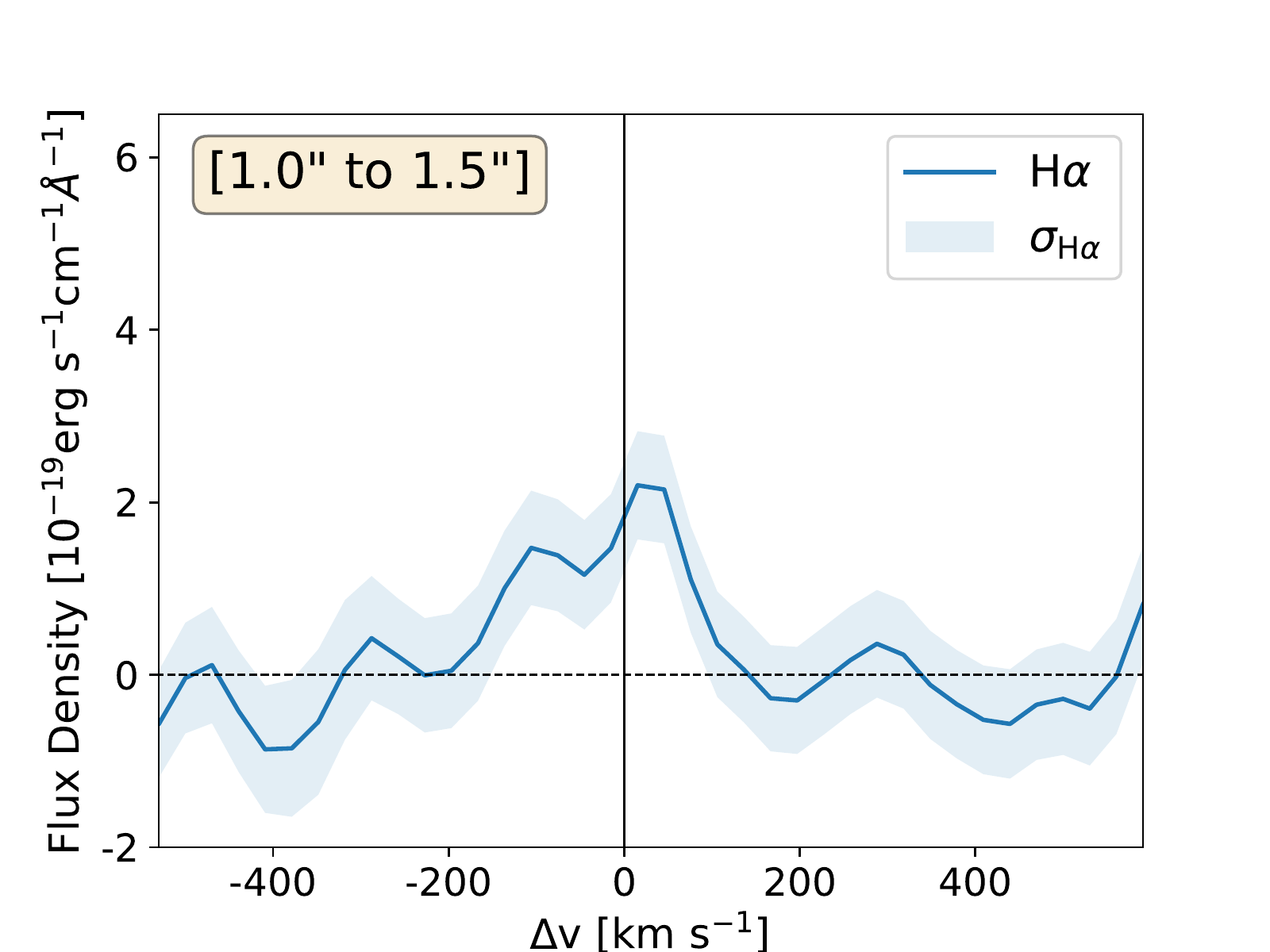}
\end{subfigure}

\begin{subfigure}{0.49\textwidth}
  \centering  
  \includegraphics[width=\textwidth]{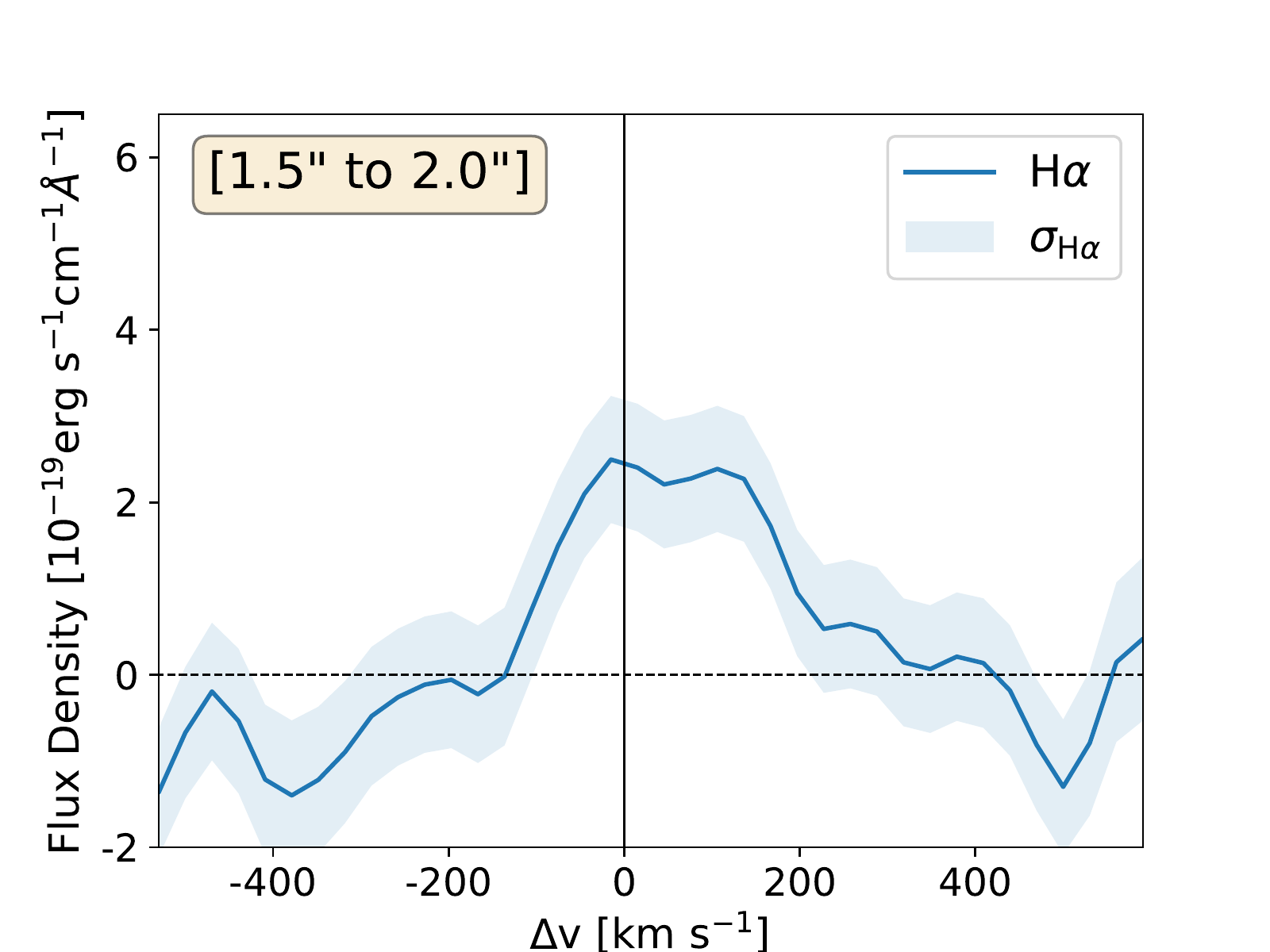}
\end{subfigure}
\hfill
\begin{subfigure}{0.49\textwidth}
    \centering  
    \includegraphics[width=\textwidth]{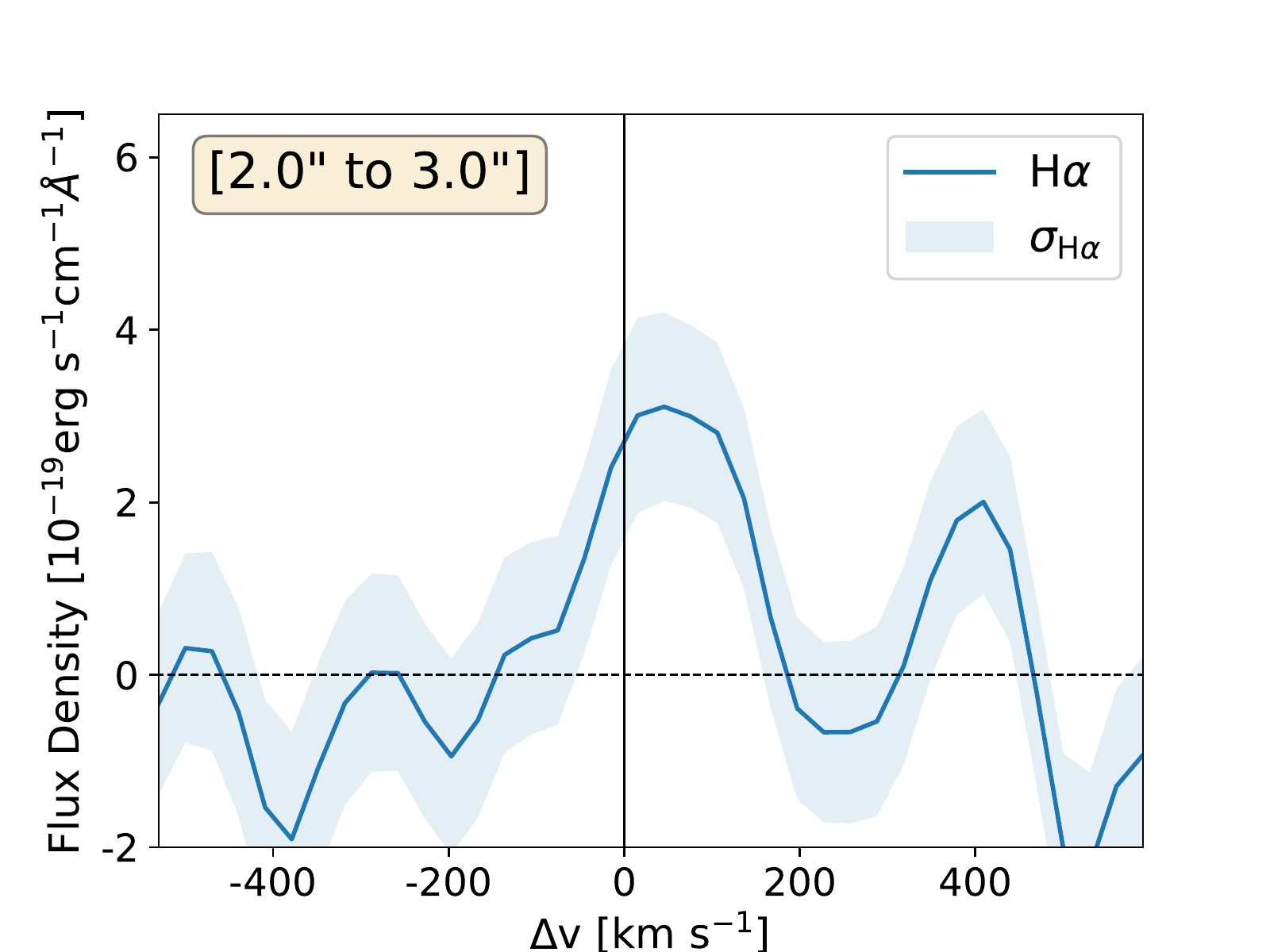}
\end{subfigure}
  
\caption[H$\alpha$ spectra as a function of distance from the quasar]{The projection of the flux density shown in Figure~\ref{fig:2DmapHa} collapsed along the spatial axis (y-axis). We have summed over 4 different windows with 2 different spatial sizes. As illustrated in Figure~\ref{app:A3} and denoted with the labels within the figure, the first 3 regions have a width of 0.5" running from 0.5" to 1", 1" to 1.5" and 1.5" to 2". The last region has a width of 1" and goes from 2" to 3". The window sizes were motivated by the peak regions of the flux seen in Figure~\ref{fig:2DmapHa} such that they encompass the brightest regions for the emission line. Note that the first window was chosen to align with the outer part of the rescaling region from the subtraction. The zero position in velocity space corresponds to the redshift of the Ly$\alpha$ nebula shown in Figure~\ref{fig:SBmapLya}. The sigma of the spectra, indicated with the shaded band, was estimated using the output of the final result of the data reduction process which has been propagated taking into account the subtraction of the quasar PSF, as explained in Sections~\ref{subsubsec:mosfire_datareduction} and~\ref{subsubsec:psfsub2D}. }
\label{fig:HaPeakShift}
\end{figure*}

Spectrally, the emission appears relatively narrow with a FWHM $<$ 300 km s$^{-1}$ with the presence of multiple emission peaks. Such complex spectral shape appears clearer in the 1D spectral profiles extracted at different positions, presented in Figure~\ref{fig:HaPeakShift}.
The most prominent double peak appears in the region closer to the quasar (0.5" to 1"), with a smaller peak near -106 km s$^{-1}$ and a slightly brighter peak near +15 km s$^{-1}$. A similar double peak structure is also present in the the two middle regions between 1" and 2" with a gradually increasing redshifted velocity to larger distances. The last spatial bin (2" to 3") present a single, broader peak around zero velocity and the possibility of a second emission peak at +400 km s$^{-1}$.
 Given the MOSFIRE spectral resolution of ~115 km s$^{-1}$, the double peaks close to zero velocity within the inner 2" are always resolved with the possible exception of the [1.5", 2"] region for which higher spectral resolution would be needed for a higher-confidence confirmation. 

% %%%%%%%%%%%%
% % PART    4%
% %%%%%%%%%%%%%%%%%%%%%%%%%%
% % COMPARISION (RESULTS)  %
% %%%%%%%%%%%%%%%%%%%%%%%%%%
\section{Comparison between the Ly\texorpdfstring{$\alpha$}{alpha} and the H\texorpdfstring{$\alpha$}{alpha} Emission}
\label{sec:comp}

In this section, we compare the Ly$\alpha$ and the H$\alpha$ emission in order to constrain both the Ly$\alpha$ emission mechanism and gas kinematics. 
For a proper comparison, we first extract the Ly$\alpha$ emission through a pseudo-slit in the KCWI data matching as close as possible the slit used in the MOSFIRE observation. 
Then, we calculate the flux ratios and compare the spectral profiles extracted within different regions in the MOSFIRE slit and the KCWI pseudo-slit.

\subsection{Ly\texorpdfstring{$\alpha$}{alpha} emission extraction in the region of the MOSFIRE slit}
\label{subsec:2Dlya}

In order to directly compare the Ly$\alpha$ and H$\alpha$ observations obtained with different techniques and instruments we ``re-observe" the Ly$\alpha$ emission using a pseudo-slit applied to the 3D KCWI data.  

More specifically, we construct a pseudo-slit with the same length, width and position angle of the MOSFIRE slit as projected on the plane of the sky as illustrated in 

Figure~\ref{fig:SBmapLya2}. 
The Ly$\alpha$ emission is then integrated along the direction of the pseudo-slit minor axis, resulting in a 2D spectral image.  
The same procedure has been used to obtain the corresponding 2D spectral image of the variance by taking into account appropriate error propagation. Finally, the astrometry has been matched using the quasar position (present in both datasets), which has been determined through a fit of the quasar PSF.

\begin{figure*}
    \centering
    \begin{subfigure}{0.49\linewidth}
      \centering
      \includegraphics[width=\textwidth]{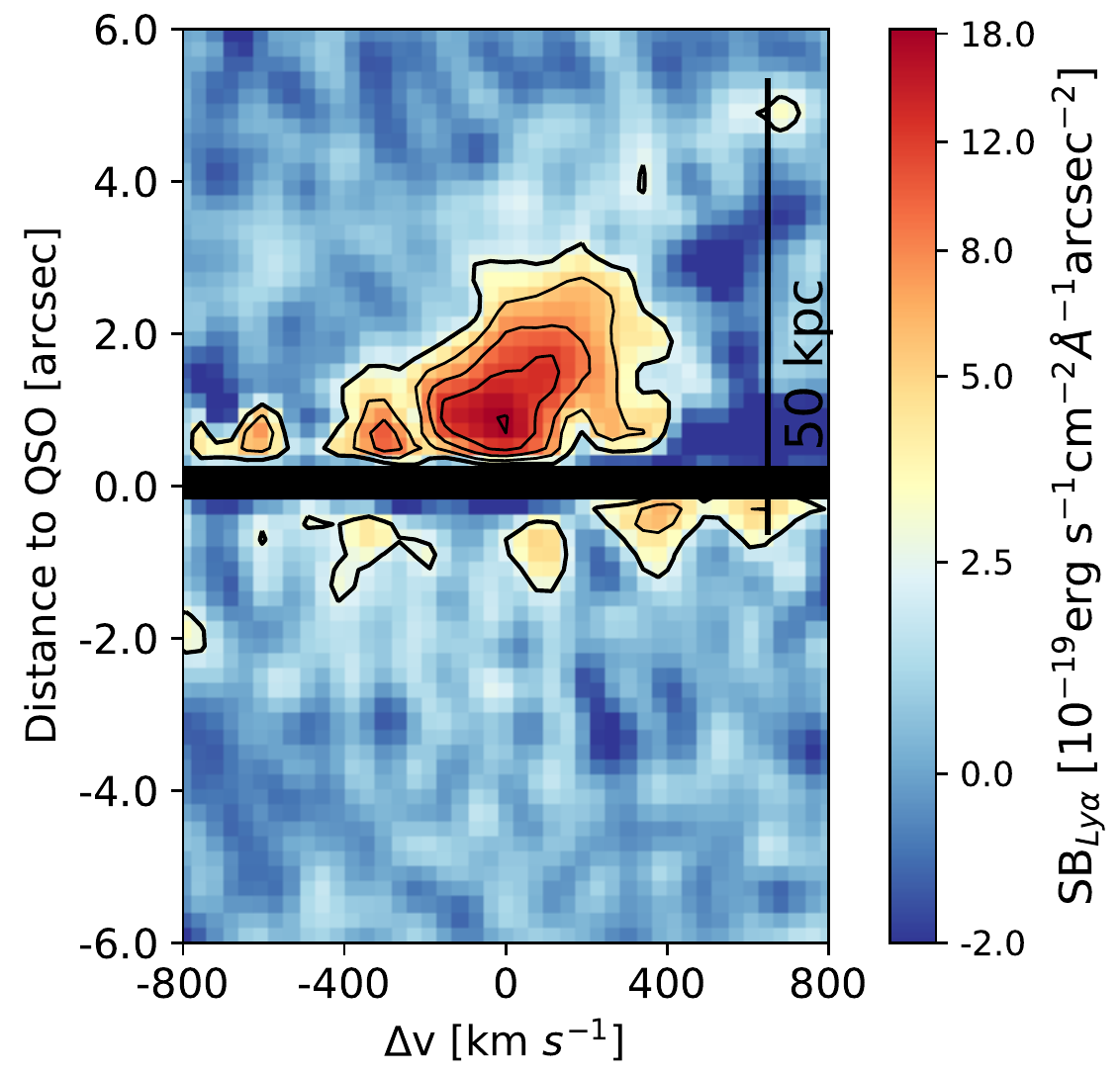}
    \end{subfigure}
    \hfill
    \begin{subfigure}{0.44\linewidth}
      \centering
      \includegraphics[width=\textwidth]{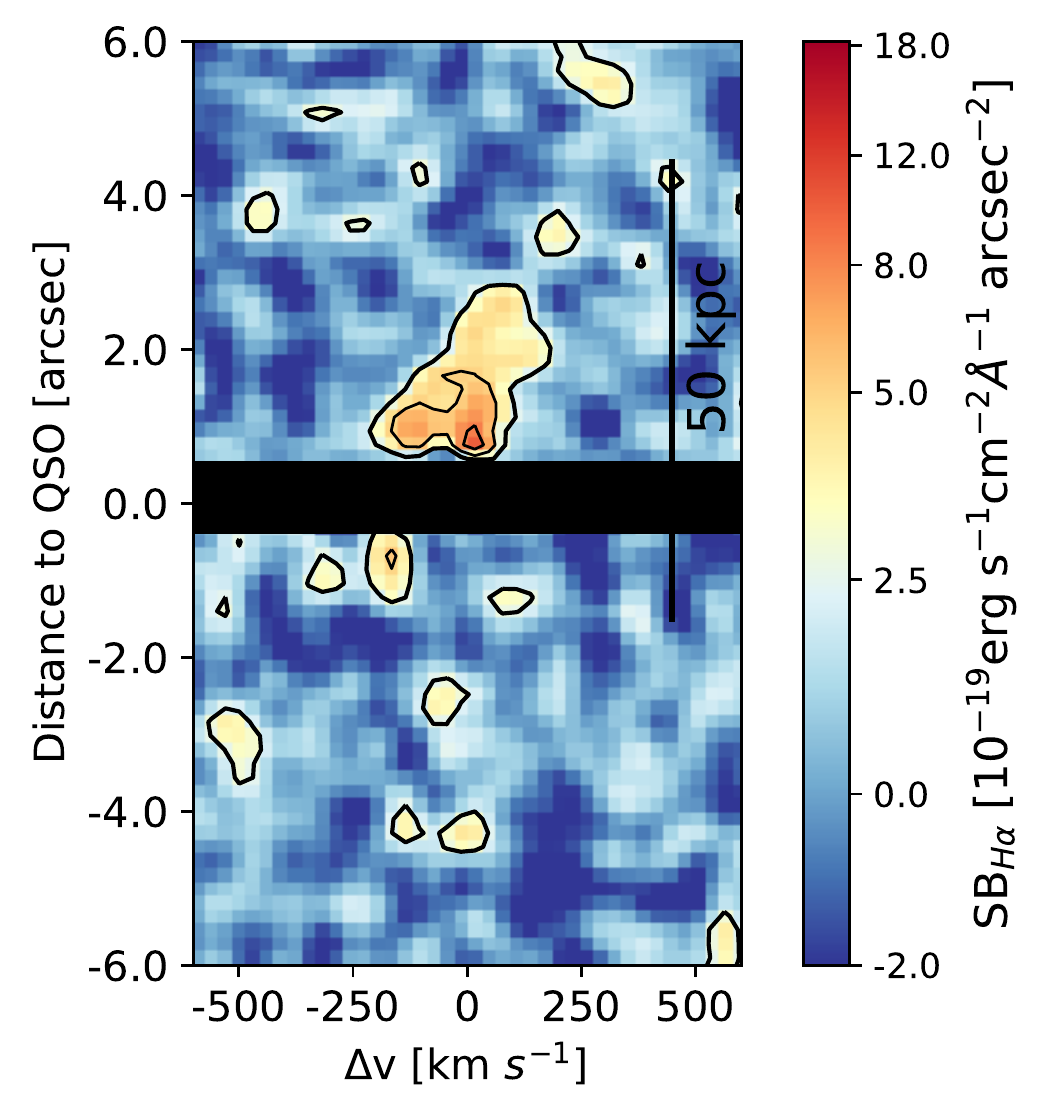} 
    \end{subfigure}
    \caption[Ly$\alpha$ flux density along the pseudo-slit]{The left panel shows the Ly$\alpha$ emission extracted from the pseudo-slit shown in Figure~\ref{fig:SBmapLya2} in units of flux density. 
    The x-axis, y-axis and the black regions represent the same as in Figure~\ref{fig:2DmapHa}. 
    The contours indicate SB levels of 2.5, 5., 8., 12. and 18. in units of 10$^{-19}$ erg s$^{-1}$cm$^{-2}\rm{\AA}^{-1}$arcsec$^{-2}$ which are indicted by the ticks on the colorbar.
    In the right panel, we adapted Figure \ref{fig:2DmapHa} to compare both emission in spatial extend and emission density. Note that the same contour levels and scales in colorbars are applied to both figures and they share share the same units of the colorbar.  }
    \label{fig:2DmapLya}
\end{figure*}

The Ly$\alpha$ emission within the pseudo-slits extends beyond 4" away from the quasar corresponding to about 35 kpc. %This is roughly twice as far as the detected H$\alpha$ emission extending to about 2". 
By comparing the H$\alpha$ flux density seen in Figure~\ref{fig:2DmapHa} and the Ly$\alpha$ flux density in Figure~\ref{fig:2DmapLya}, we observe that at each spatial location along the slit the velocity peak of the H$\alpha$ emission traces very closely that of the Ly$\alpha$ emission, with velocity shifts smaller than 100 km s$^{-1}$.
Finally, we note that, although the Ly$\alpha$ emission appears spectrally broader by a few hundred km s$^{-1}$ than H$\alpha$, the overall shape of the 2D spectral images are remarkably similar. 
A detailed analysis and quantitative comparison of the spectral profiles will be presented in 
Section~\ref{subsec:lineshapes}.

\subsection{Flux Ratios}
\label{subsec:ratios}
\begin{figure}
\centering
  \includegraphics[width=.98\linewidth]{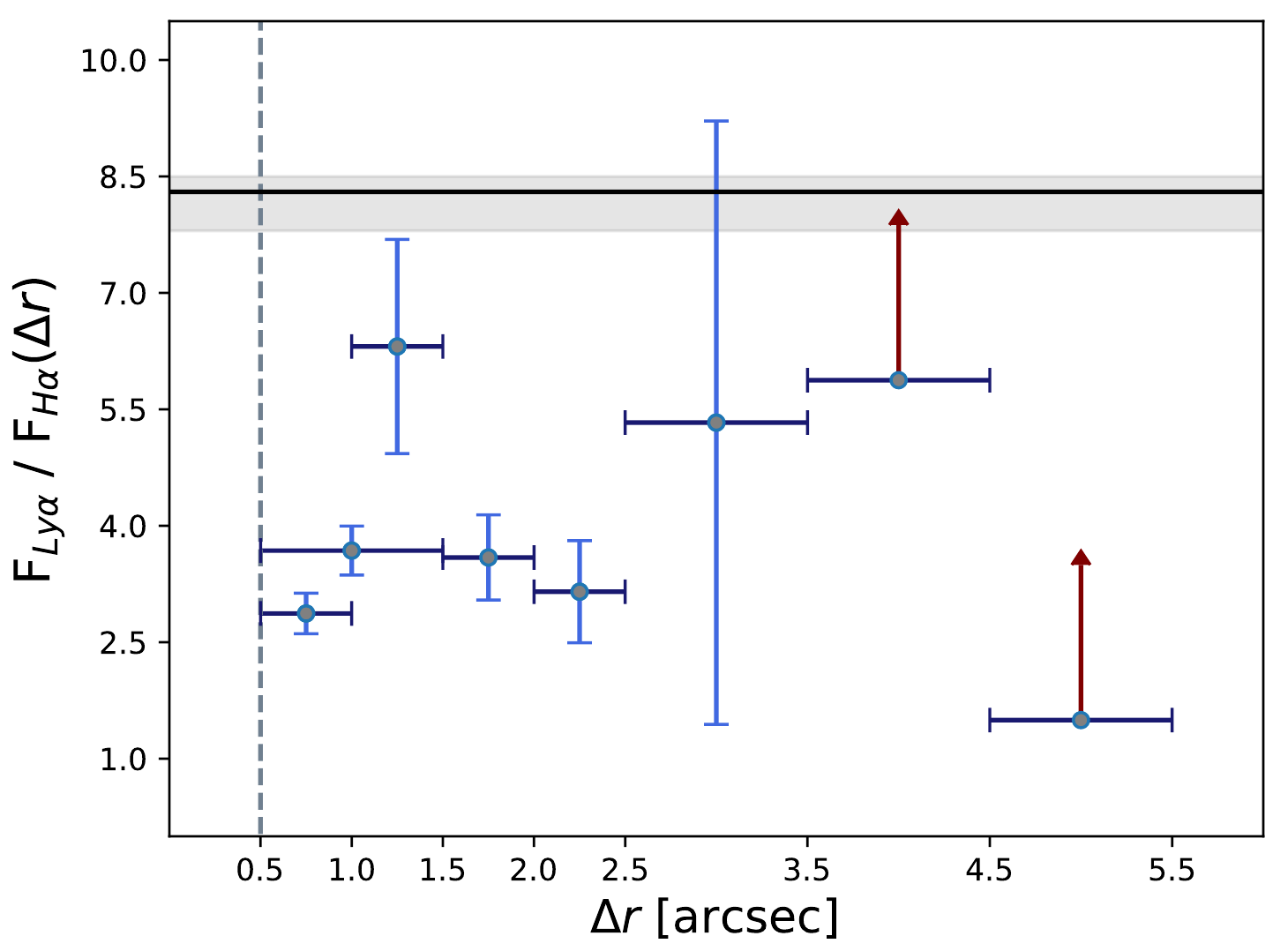}
\caption[Flux ratios as a function of distance from the quasar]{Ly$\alpha$/H$\alpha$ flux ratios obtained by integration over a fixed velocity window and within different spatial regions. The velocity windows were chosen in a way that they include the majority of the emission minimizing the flux noise as described in the text,
leading to a 500 km s$^{-1}$ window and a 1200 km s$^{-1}$ window centered at zero for H$\alpha$ and Ly$\alpha$ respectively (see Figures~\ref{fig:2DmapHa} and~\ref{fig:2DmapLya} as reference for the center positions). 
The size and position of the windows in the spatial dimension along the slit are indicated on the x-axis by the dark blue, horizontal bars (for visualization see Figure~\ref{app:A1}). The vertical light blue bars represent the error of each measurement. Also note that the outermost ratios are lower limits. \\ The grey shaded area show the expected values in the low gas density limit for the case B recombination scenario with a temperature range of T = (0.5-4)$\times$10$^4$K in absence of local Ly$\alpha$ RT effects (see Section \ref{subsec:mechanism}).
The expected ratios for both collisional excitation and ``photon-pumping" are above 100 and the corresponding lines are not reported here in the figure for clarity since all the observed flux ratios are well below these values. Such low observed flux ratios compared to the theoretical expectations suggest that recombination radiation alone, in addition to local radiative transfer effects, is fully able to explain the observed Ly$\alpha$ emission without the need to invoke additional contributions from other emission mechanisms.
}

\label{fig:windowflux}
\end{figure}

We detected a total flux within the 1" slit aperture of F$_{\tiny \textrm{H}\alpha}$=(0.95 $\pm$ 0.09) $\times$ 10$^{-17}\mathrm{erg} \ \mathrm{s}^{-1} \mathrm{cm}^{-2}$ for H$\alpha$ and a total flux of  \mbox{F$_{\tiny \textrm{Ly}\alpha}$=(3.5 $\pm$ 0.06) $\times$ 10$^{-17}\mathrm{erg} \ \mathrm{s}^{-1} \mathrm{cm}^{-2}$} within the 1" pseudo-slit aperture for Ly$\alpha$.
The flux was extracted from a spatial region of [0",5.5"] from the center and velocity windows of 500 km s$^{-1}$ and 1200 km s$^{-1}$ for H$\alpha$ and Ly$\alpha$ respectively. The regions were both centered around the peak of the Ly$\alpha$ and H$\alpha$ emission respectively, see Figure \ref{app:A1}.
Combining these measurements, we obtain a total flux ratio of: 
\begin{equation}
\frac{\textrm{F}_{\tiny \textrm{Ly}\alpha}}{\textrm{F}_{\tiny \textrm{H}\alpha}} = 3.7 \pm 0.3.    
\end{equation}
This value is consistent with the only other quasar Ly$\alpha$ nebula observed so far in H$\alpha$ emission, i.e. the Slug Nebula \citep{leibler2018detection} and indicative of a recombination origin of Ly$\alpha$ emission as discussed in detail in Section~\ref{subsec:mechanism}. 
Thanks to the high signal to noise of our H$\alpha$ detection, we can further explore the line ratio within smaller spatial regions as a function of distance from the quasar. 
In Figure~\ref{fig:windowflux}, we present the flux ratios as a function of distance from the quasar obtained within spatial windows of size $\Delta$r, as indicated by the horizontal, dark blue bars, and the same velocity windows as used above for the estimate of the integrated flux ratio. The position and size of these windows are further illustrated in Figure~\ref{app:A1}.
We find ratios with values around 3 and 2.5 for the first two regions, whereas further out the ratios increase up to a value of 5 for the third and the fourth region. For the spatial window [3.5, 4.5], we can only obtain lower limits on the flux ratio of about 6, which is still consistent with an increasing ratio towards the outer regions. Beyond 4.5" our MOSFIRE data is not deep enough for a stringent constraint of the line ratios.

\begin{figure}
\centering
  \includegraphics[width=.98\linewidth]{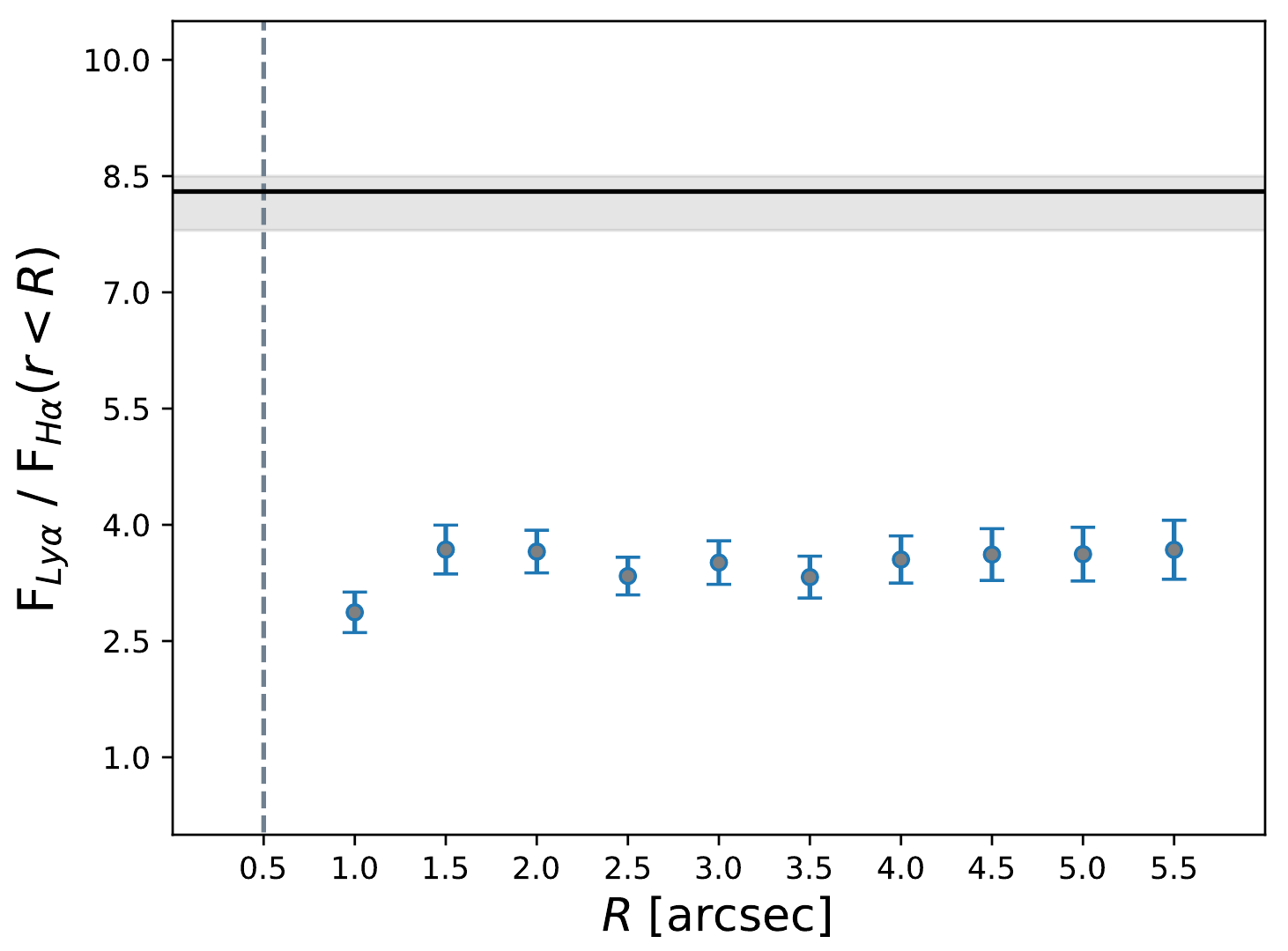}
\caption[Flux ratios in a cumulative diagram]{Each point shows the flux ratio with fluxes integrated up to the radius indicated on the x-axis. In particular, the last value at 5.5" from the QSO contains the majority of the detected H$\alpha$ and Ly$\alpha$ emission.  The black horizontal line and the grey area show the values expected for pure recombination emission without any Ly$\alpha$ absorption or radiative transfer effects as discussed in Section~\ref{sec:intro}.
As in Figure~\ref{fig:windowflux}, we are not showing here the expected line ratios from other emission mechanisms since they are all above a value of about 100 which is well above the observed flux ratios.
}
\label{fig:cumflux}
\end{figure}

In Figure~\ref{fig:cumflux}, we present the cumulative flux ratios as a function of distance from the quasar. The integration regions are illustrated in Figure~\ref{app:A2}. As expected from the integrated flux ratio measurement, the cumulative values converge to a value of about 4. Closer to the center, the ratio is instead much smaller. We discuss in Section~\ref{subsec:lineshapes} the possible physical origin of such a trend. The black horizontal line and the grey area show the values expected for pure recombination emission without any Ly$\alpha$ absorption or radiative transfer (RT) effects as discussed in Section~\ref{sec:intro}. A value within or below the grey area exclude ``scattering" or collisional excitation as possible contribution to the Ly$\alpha$ emission. We will further discuss the implication of this result in Section~\ref{sec:discussion}.

\subsection{Comparison of the Line Shapes}
\label{subsec:lineshapes}
\begin{figure*} 
\centering
  \begin{subfigure}{0.49\textwidth}
    \centering  
    \includegraphics[width=\textwidth]{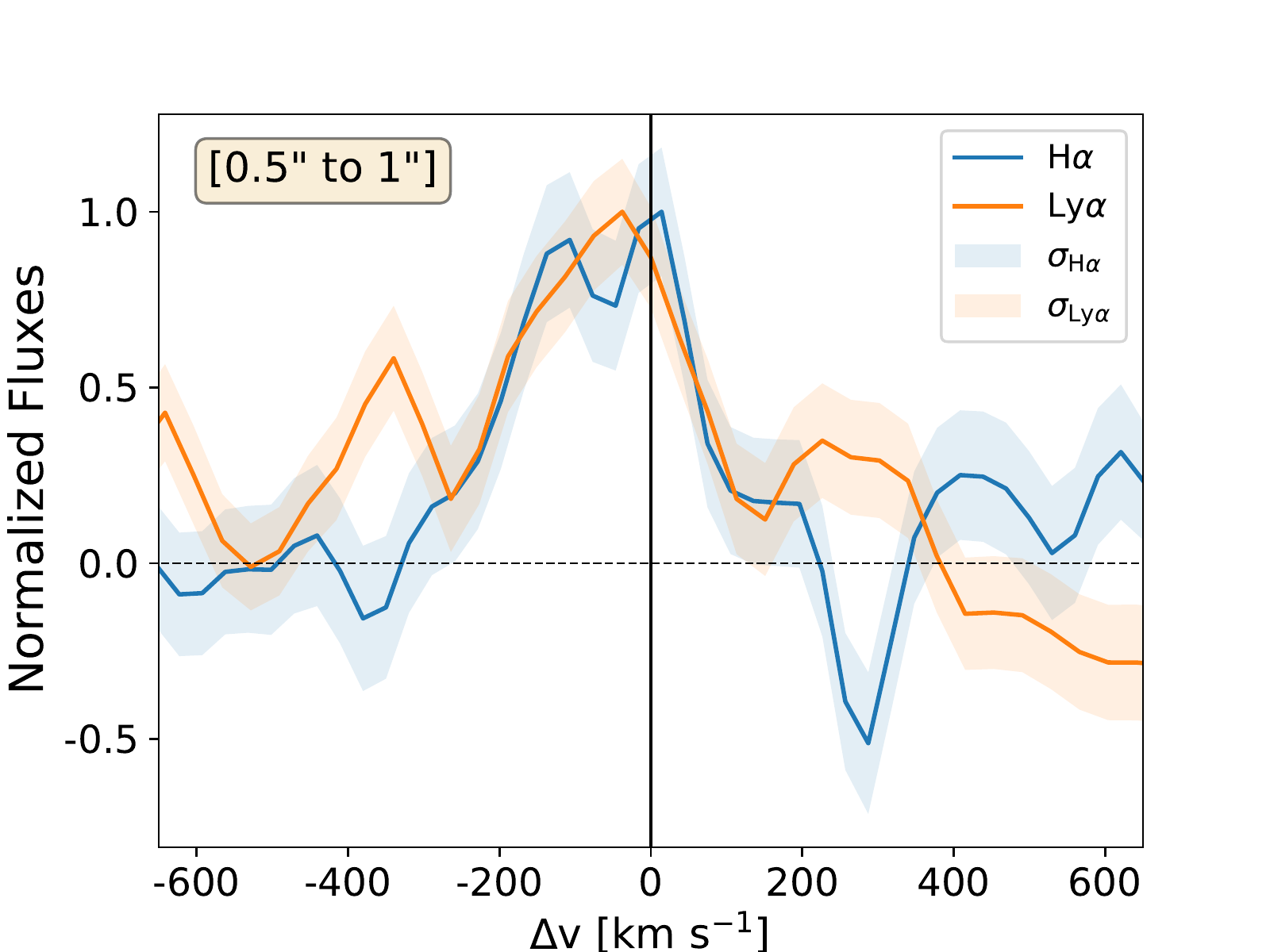}
  \end{subfigure}
  \hfill
  \begin{subfigure}{0.49\textwidth}
    \centering
    \includegraphics[width=\textwidth]{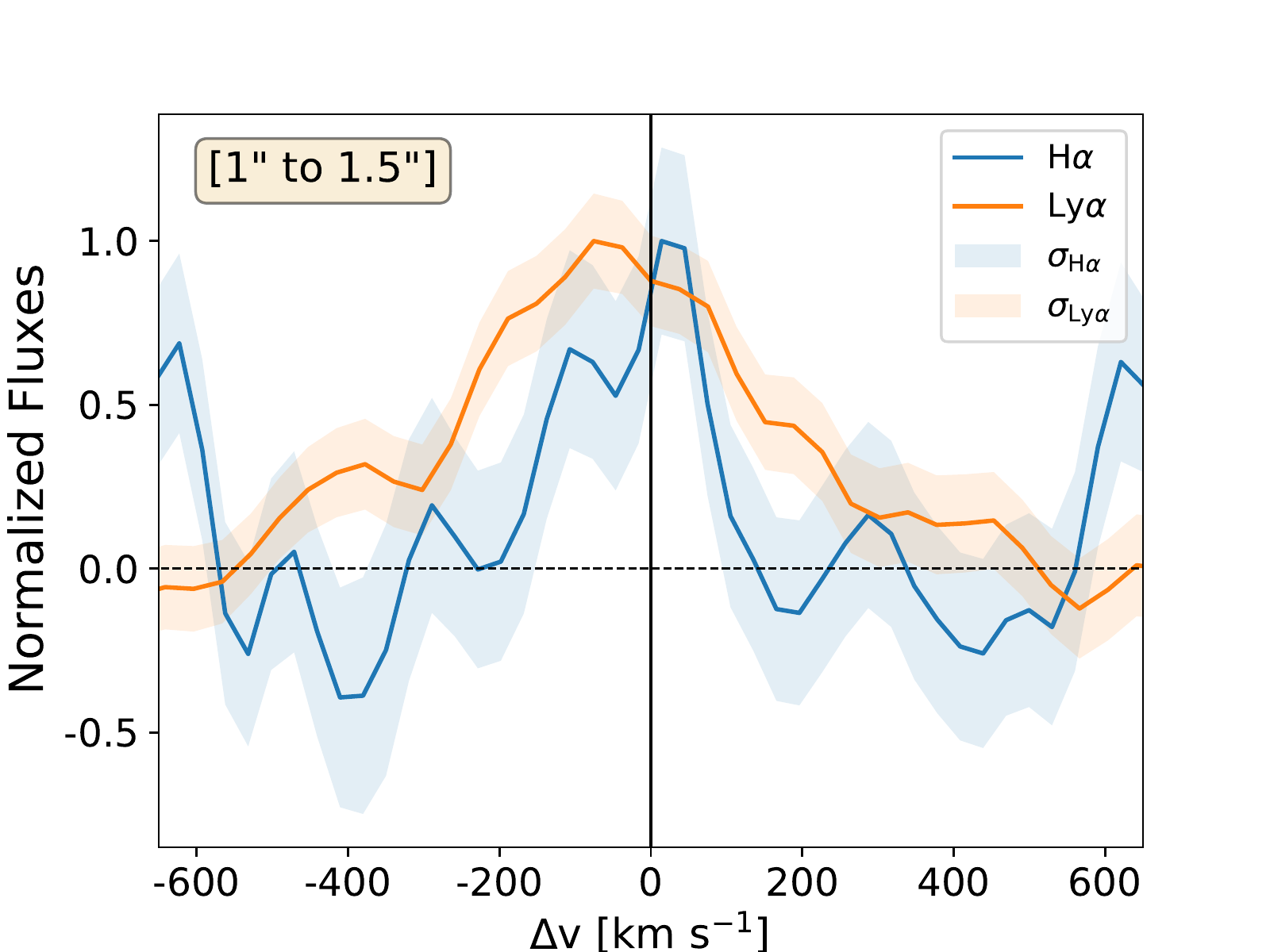}
  \end{subfigure}
  
  \begin{subfigure}[b]{0.49\textwidth}
    \centering
    \includegraphics[width=\linewidth]{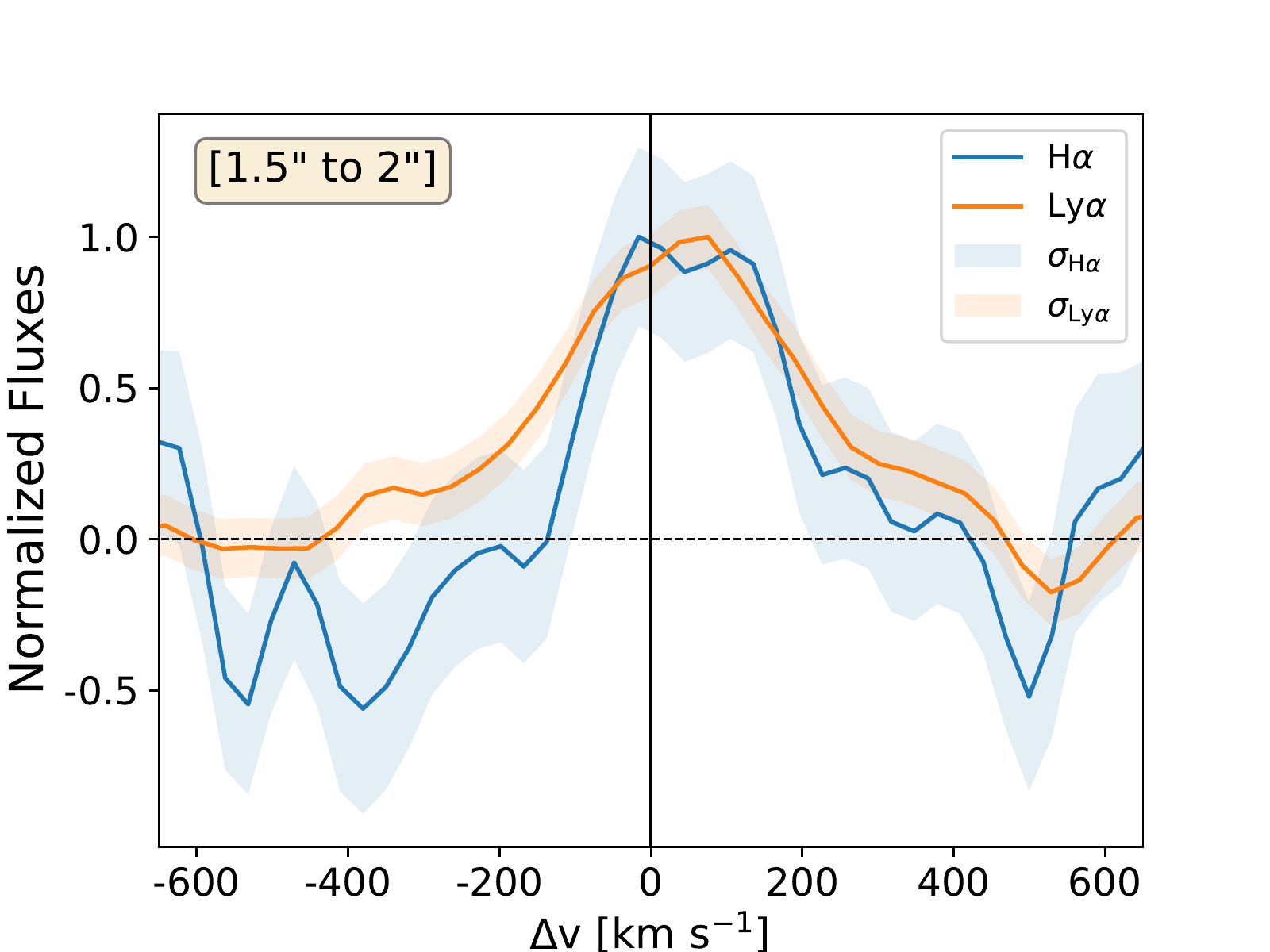}
  \end{subfigure}
  \hfill
  \begin{subfigure}[b]{0.49\textwidth}
    \centering
    \includegraphics[width=\linewidth]{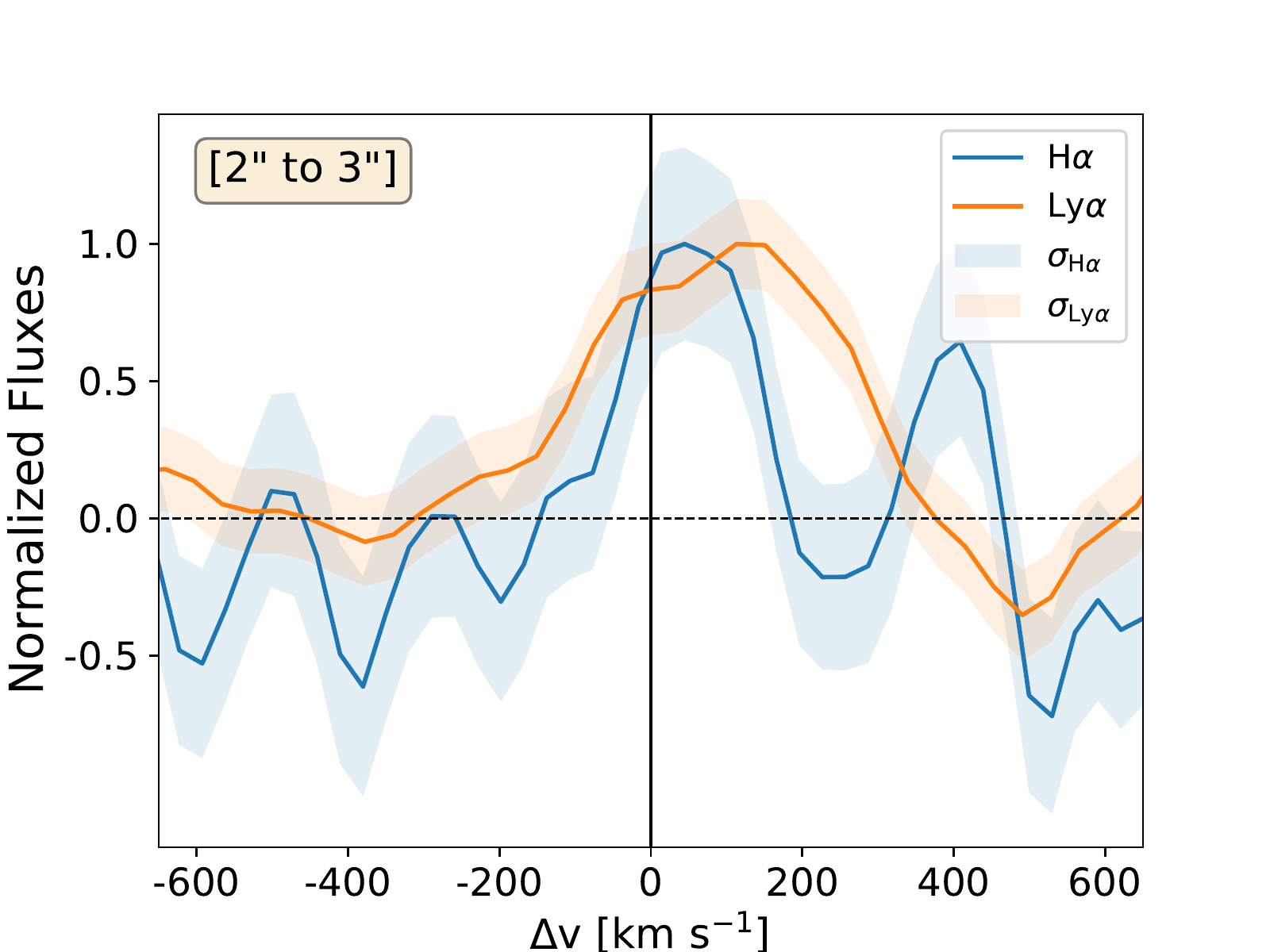}
  \end{subfigure}

\caption[Comparison between the spectra of Ly$\alpha$ to H$\alpha$ as a function of distance from the quasar]{Presented here is a comparison between the Ly$\alpha$ and H$\alpha$ spectral shape.  We projected the Ly$\alpha$ emission onto the spectral dimension as it has been done for the results shown in Figure~\ref{fig:HaPeakShift}, using also the same spatial regions. Furthermore, each of the spectra has been normalized to a peak value of 1 in order to facilitate the comparison. 
The zero velocity on the x-axis is fixed to the redshift of the Ly$\alpha$ nebula measured as the peak of the integrated Ly$\alpha$ spectrum. }
\label{fig:normCompLyaHa}
\end{figure*}

In this section, we examine the shape of the emission lines as well as their width in velocity space more closely. In order to do that, we integrate along the spatial direction the 2D spectral images within the same regions as used for the flux ratios in Figure~\ref{fig:windowflux} (illustrated in Figure~\ref{app:A3}) obtaining 1D spectral profiles.
In Figure~\ref{fig:normCompLyaHa}, we present the comparison of the 1D spectral profiles of H$\alpha$ and Ly$\alpha$ emission within the same spatial apertures. To facilitate the comparison, we have normalized each 1D spectral profile by the value of its peak emission. 

We notice again the presence of two narrow emission peaks around the zero velocity in the H$\alpha$ spectra which are not present in the corresponding Ly$\alpha$ emission. This difference cannot be explained by an instrumental effect since the spectral resolution of our KCWI data ($\sim$75 km s$^{-1}$) is higher than the resolution of our MOSFIRE observation ($\sim$115 km s$^{-1}$). Instead, this discrepancy may hint once again at the presence of radiative transfer effects which change the spectral (and spatial) shape of the Ly$\alpha$ line. These same effects are likely responsible for the broader spectral width of the Ly$\alpha$ spectral profile. Despite these differences, the Ly$\alpha$ appears in any case to be a good tracer of the overall gas kinematics  (i.e. in terms of peak position, line width and line shape, as constrained by H$\alpha$) given the good correspondence between H$\alpha$ and Ly$\alpha$ peak positions and the fact that Ly$\alpha$ is only slightly broader than H$\alpha$ (i.e. less than a factor 2).

% %%%%%%%%%%%
% PART    4%
% %%%%%%%%%%%%%
% DISCUSSION %
% %%%%%%%%%%%%%

\section{Discussion}
\label{sec:discussion}

The detection of circum-QSO H$\alpha$ emission at high signal-to-noise level from a quasar Ly$\alpha$ nebula at $z=2.257 \pm 0.005$ gives us the unique opportunity to constrain both the Ly$\alpha$ emission mechanism (and thus the gas density distribution) and the gas kinematics as discussed in detail below. Before discussing the implication of our results, we will review and discuss the Ly$\alpha$ emission mechanism and calculate their contribution to the expected Ly$\alpha$/H$\alpha$ flux ratios in the context of the CGM and IGM. 

\subsection{Ly\texorpdfstring{$\alpha$}{alpha} emission mechanisms and Ly\texorpdfstring{$\alpha$}{alpha}/H\texorpdfstring{$\alpha$}{alpha} ratios}
\label{subsec:mechanism}
There are three physical processes that can produce a Ly$\alpha$ photon; \textit{(i)} recombination radiation arising from photo-ionized hydrogen;\textit{(ii)} ``continuum-pumping'' in which Ly$\alpha$ photons are produced, e.g., in the quasar Broad-Line region and are scattered into our line of sight; and \textit{(iii)} Ly$\alpha$ "cooling radiation", in which the emission is generated through collisional excitation of a neutral hydrogen atom by a free electron (see also Section~\ref{sec:intro} for further details). 
When the emitting gas is photo-ionized by an external source, e.g. in the case of a quasar illuminating the CGM and the IGM, the radiation produced by process \textit{(i)} is usually referred to as ``flourescent'' in the literature, e.g. \cite{gould1996imaging}, \cite{cantalupo2005fluorescent}, \cite{cantalupo2012detection}, \cite{borisova2016ubiquitous}. This is because it can be described as a re-processing of external ionizing photons, with lambda $<$ 912 $\AA$, into photons with lower energy and longer wavelength (e.g. Ly$\alpha$ or H$\alpha$).
This is not the case for process \textit{(ii)} nor for process \textit{(iii)}, which is not initiated by radiation. \par
The actual number of Ly$\alpha$ or H$\alpha$ photons produced on average for each recombination event in a gas cloud depends on the fate of the Lyman continuum and the resonant Lyman series photons (excluding Ly$\alpha$), which are also generated during atomic recombination. Two extreme cases are usually assumed in the literature (starting from \cite{baker1938physical}) : `case A', for which all Lyman continuum and Lyman series line photons completely escape the emitting region without interacting with the gas, and the opposite situation, referred as `case B'. We consider that in `case B' Ly$\alpha$ photons will eventually escape the emitting region through a series of absorption and re-emission events (``scatterings'') since they cannot be ``transformed" to any other line emission during these events for the densities relevant to the CGM/IGM (and also ISM). In most realistic situations, neither of the two cases provide a good representation of the actual recombination rates, e.g. \cite{cantalupo2008mapping}, however they are useful to bracket the possible ranges of interest. \par
In the literature, the Ly$\alpha$/H$\alpha$ flux ratios for recombination radiation is often quoted to be in the 
range 8.1 - 11.6 for temperatures between 5$\times10^3$ K and 2$\times10^4$ K and electron density in the range $n_e = 10^{2}$ - $10^{4}$ cm$^{−3}$
from the work of \cite{hummer1987recombination}, e.g. \cite{hayes2015lyman}. A typical value of Ly$\alpha$/H$\alpha=8.7$ (corresponding
to a temperature of $10^4$ K and $n_e=350$ cm$^{-3}$) is often used. In particular, these values have been derived considering the typical densities of nebulae in the ISM of galaxies and therefore could be more representative of line ratios for galaxies rather than the CGM or IGM. Above a critical density, atomic collisions are responsible for the mixing of orbital angular momentum levels (``$l$-mixing collisions"). In particular, higher densities typically correspond to an increase in Ly$\alpha$ production rate due to collisional transfer between the 2s and 2p states. This is due to the much longer lifetime of the metastable 2s state (which can only decay to 1s through two-photon emission) compared to 2p.  Collisions have a smaller impact on H$\alpha$ emissivities since the 3s, 3p and 3d are not metastable. Therefore, higher densities typically result in larger Ly$\alpha$/H$\alpha$ flux ratios at a given temperature. 

Since typical CGM and IGM densities are expected to be much lower than $n_e = 10^{2}$ cm$^{−3}$ we have re-estimated the Ly$\alpha$/H$\alpha$ flux ratios in the low density limit (i.e., ignoring $l$-mixing collisions) using the effective recombination coefficients obtained by \cite{martin1988hydrogenic} at T$<2\times10^4$ K and by \cite{pengelly1964recombination} at higher temperatures for both Case A and Case B. In particular, we have used the tabulated $\alpha^{(c)}_{2p}$ values for Ly$\alpha$ and the sum of $\alpha^{(c)}_{3s}$,  $\alpha^{(c)}_{3p}$, and $\alpha^{(c)}_{3d}$ for the H$\alpha$ effective recombination coefficient, multiplying $\alpha^{(c)}_{3p}$ by the 3p to 2s branching ratio of about 0.1183 in Case A only, in order to take into account the production of Ly$\beta$ photons. 
As a result, we have obtained Ly$\alpha$/H$\alpha$ flux ratios ranging between 7.8 and 8.5 for temperatures between 5$\times10^3$ K and 4$\times10^4$ K in Case B with a value of 8.3 at T$=2\times10^4$ K. 
Considering Case A, we obtain instead a flux ratio in the range 10.5 to 13.7 in the temperature range between 5$\times10^3$ K and 4$\times10^4$ K, with a value of 12.7 at T=$2\times10^4$ K. The difference between these two cases is almost entirely due to the assumed fate of the Ly$\beta$ photons originated from the 3p state. 
Indeed, in Case B, all Ly$\beta$ photons are eventually converted into H$\alpha$ photons, resulting in a significant boost of the H$\alpha$ flux. On the other hand, the Ly$\alpha$ flux is boosted only modestly, as a consequence of (only partial) re-conversion of the (less numerous) Ly$\gamma$ photons (and higher-n Lyman series). For the precise values, we refer the reader to the tables in \cite{pengelly1964recombination}.

We assume that the flux ratios provided by the Case B are the most relevant in our situation because of the high Ly$\beta$ absorption cross section (about 19\% of the Ly$\alpha$ absorption cross section at line center). Indeed, Ly$\beta$ opacity could be relatively high even for modest \ion{H}{I} column densities which are possible even for the highly ionized gas around quasars. Note that, although we consider case B to be more likely, assuming case A would only reinforce our conclusions.   

The other two processes which could generate Ly$\alpha$ photons are instead expected to produce either negligible (for ``collisional excitation'') or no H$\alpha$ emission at all (``continuum-pumping"). In particular, we have used the fits as a function of temperature of the effective collision strength provided by \cite{giovanardi1987numerical} to calculate the collisional excitation rates producing Ly$\alpha$ and H$\alpha$ photons including the contribution from levels up to n=4 in the low density limit as above. The expected  Ly$\alpha$/H$\alpha$ flux ratios from collisional excitations range from values of ~100 to ~120 for temperatures between $1.5\times10^4$ K and 4$\times10^4$ K in Case B\footnote{Temperatures lower than $1.5\times10^4$ K do not produce Ly$\alpha$ photons efficiently. Therefore lower temperatures could result in lower line ratios, but would not produce sufficiently bright emission in first place.}. Even higher values of the flux ratios are obtained in Case A, for the same reasons as in the discussion above. 
H$\alpha$ photons cannot be produced by ``continuum-pumping" 
at the CGM (and even ISM) densities. 
Thus, if there is any contribution to the Ly$\alpha$ emission from mechanisms different than recombination we should expect a flux ratio Ly$\alpha$/H$\alpha$ $>8.5$. Any increase of this ratio with respect to the expectations from recombination radiation could be used in principle to estimate the relative contribution of the different emission mechanisms.

Are there, on the other hand, any processes or effects that could instead reduce the measured Ly$\alpha$/H$\alpha$ flux ratios with respect to the recombination case? Dust absorption is expected to be much more significant for Ly$\alpha$ than for H$\alpha$ given the bluer Ly$\alpha$ wavelength reducing the line ratios. The actual effect of dust depends on two factors: i) the amount and spatial distribution of the dust, ii) the number of scatterings which Ly$\alpha$ suffers before escaping the medium. Both these factors are very difficult to estimate but they are in general expected to be very relevant in the ISM of galaxies, reducing the observed Ly$\alpha$ emission (and thus the Ly$\alpha$/H$\alpha$ flux ratios) by about one order of magnitude \citep[e.g.][]{hayes2010escape}. Although there are no direct observational constraints on the dust content of the CGM and IGM, the smaller metallicity with respect to the ISM as commonly suggested by cosmological simulations \citep[e.g.][]{wright2021revealing, mitchell2022baryonic}, might indicate also a much smaller dust content.
Moreover, the strong UV, FUV and X-ray flux produced by the quasar might also contribute to dust evaporation. In summary dust attenuation may contribute to the smaller line ratios, however we do not expect the same reduction as observed for Ly$\alpha$ emitting galaxies. 

Even in absence of dust, the local scattering of Ly$\alpha$ photons (``radiative transfer effects") is expected to produce a reduction in the Ly$\alpha$ Surface Brightness while the H$\alpha$ Surface Brightness (SB) will be unaffected. This is due to both scatterings within the medium and to intervening material at large distances depending on its velocity field relative to the emitting region. While in the first case the total Ly$\alpha$ emission could be recovered, e.g., by integrating within a larger aperture, in the case of intervening material, the Ly$\alpha$ photons are effectively lost from our line of sight (similar to what happens in the case of absorption lines in quasar spectra). The effective decrease in Ly$\alpha$ SB due to these radiative transfer effects will depend on the detailed geometrical and kinematical properties of the medium and its large scale environment, in addition to the local ionization parameters. Early numerical simulations of fluorescent emission showed that such effects can account for large SB variations
\citep[see][]{cantalupo2005fluorescent}, however more detailed numerical studies are needed for a proper estimation. If Ly$\alpha$ radiative transfer effects are important, we expect that comparison between Ly$\alpha$ and H$\alpha$ fluxes made on fixed apertures, as in our case, will result in  Ly$\alpha$/H$\alpha$ flux ratios lower than the recombination ratio discussed above. 

\subsection{Implications of the measured Ly\texorpdfstring{$\alpha$}{alpha}/H\texorpdfstring{$\alpha$}{alpha} flux ratio} 

The measured Ly$\alpha$/H$\alpha$ flux ratio in our case is about 4 when integrated within the slit aperture of our MOSFIRE observations, i.e. about a factor of 2 smaller than the Case B recombination ratio (and about a factor of 3 smaller than the Case A recombination ratio). The total fluxes of both Ly$\alpha$ and H$\alpha$ are dominated by the regions closer to the QSO. From the discussion above, the first implication of our results is the presence of Ly$\alpha$ radiative transfer effects in the (inner) CGM of J0010+06 quasar. This is not surprising, given that gas becomes opaque to Ly$\alpha$ photons at very modest neutral column densities, e.g. about 10$^{14}$ cm$^{-2}$, which are typically observed even in proximity of bright quasars. The observed differences between the Ly$\alpha$ and H$\alpha$ line shapes also suggest the presence of radiative transfer effects. However, the fact that these line shapes are much more similar (in terms of, e.g., line width, symmetry of the line and the wavelength location of the peak) than observed in galaxies, suggest smaller \ion{H}{I} column densities and thus weaker radiative transfer effects in our case. Such smaller \ion{H}{I} column densities are expected given the strong quasar ionizing radiation and the lower densities of the CGM/IGM with respect to the ISM case. A detailed description of strong radiative transfer typically observed in the ISM of galaxies can be found in \citep{dijkstra2006lyalpha}. 
We note that our results appear to indicate a larger Ly$\alpha$/H$\alpha$ flux ratio, possibly approaching the case B recombination value, at larger distances from the quasar. This could be an indication that (more) neutral gas or dust is present in the inner CGM (despite the closer proximity to the quasar\footnote{We do not expect more neutral gas if the density increases with decreasing distance stronger then d$^{-2}$.}). Alternatively, it could simply reflect a geometrical factor, i.e. the larger path-length traveled by the Ly$\alpha$ photons along our line-of-sight in order to escape the quasar host halo.\par
In all cases, the measured H$\alpha$ flux, which can only be produced by recombinations as discussed above, could be used to recover the expected Ly$\alpha$ SB produced by recombinations in absence of scattering effects. We will refer to the Ly$\alpha$ emission produced through recombination as the ``intrinsic" fluorescent Ly$\alpha$ SB. Our results imply a correction factor of up to about 2 to convert the observed Ly$\alpha$ SB into the intrinsic fluorescent Ly$\alpha$ SB in the region covered by the MOSFIRE slit (in the likely situation approximated by Case B). A similar correction factor was also obtained for the Slug Nebula as observed through the MOSFIRE slit. 
Assuming that such correction factors apply also to other regions of the nebulae not covered by the MOSFIRE slits and to the other bright quasar Ly$\alpha$ nebula allows us to translate the observed Ly$\alpha$ SB into their ``intrinsic" counterparts. In turn, the ``intrinsic" fluorescent Ly$\alpha$ SB can be used to estimate gas densities, gas clumpiness and other properties of the CGM as discussed, e.g., in \cite{cantalupo2014cosmic}, \cite{cantalupo2019large}, \cite{pezzulli2019high}, confirming the need for broad gas density distributions or very ``clumpy" and dense medium (i.e., with log-normal $\sigma>2$ if a log-normal density distribution is assumed) in the CGM of bright quasars.
%%%%
In addition, taking into account the analytical model of \cite{pezzulli2019high}, the fact that recombination radiation is the dominant mechanism would imply that close to 100\% of the expected baryons are retained within the virial radius of QSO-hosting halos; it would also indicate that either the bright
quasars live in much more massive halos than currently expected from clustering measurements or that cold emitting gas in the CGM is out of pressure equilibrium with the ambient hotter halo gas (see \citealt{pezzulli2019high} for details). The latter situation is verified in particular if a broad gas density distribution is responsible for the Ly$\alpha$ emission in the CGM. Such broad density distributions could be generated by hydrodynamical effects (e.g. \cite{vossberg2019density}, Vossberg et al., in prep.). Finally, another possibility could be that the emitting gas did not have time to reach pressure equilibrium yet, short after being photo-ionized (and heated) by the QSO radiation, depending on the QSO "age" and/or the size of the cold gas emitting structures (e.g. \cite{pezzulli2019high}, Sarpas et al., in prep.).
As shown in Figure \ref{fig:windowflux}, our results could also indicate that the correction factors decreases at larger distances from the quasar. If this trend will be confirmed by additional data, this could also imply a ``steepening" of the intrinsic Ly$\alpha$ SB profiles with respect to the observed Ly$\alpha$ SB profile which the models should take into account for a proper comparison (as adressed in Section 6.2 of \cite{pezzulli2019high} .

\subsection{Gas Kinematics}
\label{subsec:kinematics}

In this paragraph, we discuss the kinematics of the gas. We would like to stress again that Ly$\alpha$ is in principle not an ideal tracer of kinematics due to its resonant nature and hence a detailed examination in the light of a non-resonant emission line, as has been done in this work, is crucial.
The first thing we notice is that none of the H$\alpha$ line profiles has the shape of a single Gaussian, suggesting a more complex origin than a simple gas cloud, e.g. the presence of multiple structures in projection as also observed for the Slug Nebula \citep[e.g.][]{cantalupo2019large}.
In order to study the gas kinematics in more detail we focus on two quantities: i) the shift of the line peaks relative to each other, ii) the nebula centroid as a function of distance from the quasar, and iii) the H$\alpha$ line width, which could be related to the overall kinematics along our line of sight or gas turbulence. \par
We relate our results to two different scenarios of strong bulk motions: quasar outflows and disk-like rotation. In case of a strong gas outflows, we would expect gas outflow velocities v$_s$ $>$ 1000 km s$^{-1}$ near the QSO, as argued in \cite{arrigoni2019qso} %more refs here
and modeled in \cite{allen2008mappings}, \cite{carniani2015ionised} and \cite{valentini2021host}. 
However, some theoretical studies suggest that outflow velocities could diminish with distance from the central source, see e.g. \cite{faucher2012physics}, \cite{richings2018origin} and \cite{menci2019outflows}.

Disk-like rotation requires a monotonically increasing velocity with increasing distance (along the kinematic major axis, which in our case may or may not coincide with the direction of the MOSFIRE slit), and a flattening at larger distances as modeled and in detailed described in \cite{martin2019multi} including a discussion of observational criteria, and as has been considered in \cite{martin2016newly} and \cite{arrigoni2018inspiraling}.
We want to emphasize however then these models of classical galaxy rotation curves assume thin disks on kpc or even sub-kpc scales, which may not be applicable for the nebula studied in this work. \par
%providing information on temperature or turbulence of the gas. 
In particular,  we have obtained the H$\alpha$ line profiles in 4 different distance bins from the QSO, as shown in Figure~\ref{fig:HaPeakShift}. The primary (brightest) peaks move very little as a function of distance to the quasar: we measure no velocity shift moving from bin [0.5", 1"] to [1", 1.5"], a shift of about 30 km s$^{-1}$ and 60 km s$^{-1}$ moving from [1", 1.5"] to [1.5", 2"] and [1.5", 2"] to [2", 3"] respectively. Given our spectral resolution of 115 km s$^{-1}$ these shifts can be considered insignificant. 
All peaks are centered close to the Ly$\alpha$ nebula redshift: we calculate velocity shifts relative to the systemic for the H$\alpha$ line of +14 km s$^{-1}$, +14 km s$^{-1}$, -16 km s$^{-1}$ and +45 km s$^{-1}$ and the Ly$\alpha$ line of -38 km s$^{-1}$, -75 km s$^{-1}$, +75 km s$^{-1}$ and +113 km s$^{-1}$ for the bins [0.5", 1"], [1", 1.5"], [1.5", 2"] and [2", 3"] respectively.
Hence, bulk motions of the gas on large scales along the direction parallel to our line of sight, typically observed for both outflows and disk-like rotation, are disfavoured. %\par
Perhaps even more than by the absence of significant spatial gradients, disk-like rotation is disfavoured by the presence of multiple components, as the rotation models require one connected structure and smooth velocity gradients, likewise described in \citep{martin2015giant}, \citep{martin2016newly} and observed in \citep{arrigoni2018inspiraling}. 
Strong outflows, on the other hand, are disfavoured by the small shift compared to systemic and mostly by the small FWHM of the H$\alpha$ line: we measure FWHM of about 273 km s$^{-1}$, 243 km s$^{-1}$, 303 km s$^{-1}$ and 212 km s$^{-1}$ for the bins [0.5", 1"], [1", 1.5"], [1.5", 2"] and [2", 3"] respectively. Note that these values encompass both peaks. Correcting for the effect of instrumental broadening, given our resolution of 115 km s$^{-1}$, would further decrease this value. Moreover, the H$\alpha$ profiles show multiple peaks which imply that the FWHM of each individual kinematic component is even narrower.
An exception to the above conclusion is posed by the second peak in the last bin [2", 3"] located at about +400 km s$^{-1}$, which could be either an inflow or outflow depending on its location along our line of sight. 
We would like to stress that we cannot draw definitive conclusions on the gas kinematic given our current spatial resolution. We can however notice that none of the observed components seem to exceed the expected outflow escape velocity,  expected around $\sim$500 km s$^{-1}$ for typical host halos of quasars at this redshift. This implies that if an outflow is in place, most energy and metal content should remain in the CGM and that inflows could prevail in determining the CGM gas content.\par
Despite the slightly larger width of the Ly$\alpha$ spectral profiles, they give nonetheless a good representation of the overall kinematics on large scales. We find FWHM of about 302 km s$^{-1}$, 415 km s$^{-1}$, 377 km s$^{-1}$ and 415  km s$^{-1}$ for the bins [0.5", 1"], [1", 1.5"], [1.5", 2"] and [2", 3"] respectively. However, detailed features such as the double peaks (as seen in the H$\alpha$ profiles) corresponding to multiple velocity components are likely washed out by radiative transfer effects. In any case, except for the outermost region (i.e., [2" to 3"]) for which the SNR is modest, the Ly$\alpha$ peak velocity traces fairly well the average velocity of the H$\alpha$ double-peaks, suggesting that local radiative transfer effects are only ``smoothing" the profiles rather than red-shifting them as in the case of Ly$\alpha$ emitting galaxies. Together with the information on the line ratio, this could imply that the Ly$\alpha$ photons experience only a modest number of scattering events in a low-opacity medium and therefore they mostly diffuse spatially rather than spectrally. This is the opposite situation with respect to a high-Ly$\alpha$-opacity medium such as the ISM for which the most effective escape route for the Ly$\alpha$ photons is to diffuse away from the line center through a large number (of the order of $10^{4}$ or larger) of close-by ``scattering" events.

Future H$\alpha$ Integral-Field Studies and analysis encompassing a larger area of single nebulae or a larger sample of nebulae will provide more information about the detailed kinematics of the gas which can then be compared to numerical models of fluorescent H$\alpha$ emission in a cosmological context. Such a comparison with numerical models (which are expected to be much more reliable for non-resonant recombination lines such as H$\alpha$ with respect to Ly$\alpha$) will give us the opportunity to constrain both the quasar halo mass and the origin of the cold CGM component in terms, e.g., of cosmological inflows or quasar/galactic outflows.

\section{Summary} 
\label{sec:conclusion}

The recent discovery of ubiquitous giant Ly$\alpha$ nebulae around quasars at $z>2$ has opened up the possibility to directly study, in emission, the CGM and IGM. 
However, the resonant nature of the Ly$\alpha$ line and its different possible emission mechanisms complicate the constraining of the kinematics and physical properties of the detected gas, in terms, e.g. of its density.  \par
We have presented here the results of a pilot observational survey which overcame these limitations by combining specifically designed KCWI Ly$\alpha$ emission and MOSFIRE H$\alpha$ emission observations of quasar nebulae at $z\sim2.2$. Because H$\alpha$ emission is shifted into the IR at z$>2$, the presence of bright sky lines limits the possibility of detecting it for the vast majority of previously detected quasar Ly$\alpha$ nebulae. \par
For this reason, we have first searched with KCWI for extended Ly$\alpha$ emission around three bright SDSS quasars with estimated systemic redshifts (from SDSS pipeline) within the optimal range $2.25<z<2.27$ which is free from bright IR sky lines for H$\alpha$. 
Among the three fields, which all show Ly$\alpha$ emission extending on scales larger than 40 physical kpc, we have selected the brightest and most extended nebula with a confirmed nebular Ly$\alpha$ redshift in the range $2.25<z<2.27$ (J0010) for follow-up H$\alpha$ spectroscopic observation with MOSFIRE .  \par
 Within the MOSFIRE slit on J0010, we detected H$\alpha$ emission extending up to about 20 physical kpc from the quasar with a total H$\alpha$ flux of F$_{\tiny \textrm{H}\alpha}$=(7.33 $\pm$ 0.04) $\times$ 10$^{-18}\fcgs$. By extracting the Ly$\alpha$ flux from the same spatial aperture, we calculated a line flux ratio of F$_{\tiny \textrm{Ly}\alpha}$/F$_{\tiny \textrm{H}\alpha}$=3.7 $\pm$ 0.3, consistently with the results obtained for the Slug Nebula at z$=2.275$ (whose H$\alpha$ measurement is affected, however, by the presence of a sky line). This relatively low line ratio compared to our expectations from different emission mechanisms (see Section 6.2) suggest that recombination radiation is the dominant Ly$\alpha$ mechanisms confirming the need for broad gas density distributions or very ``clumpy" and dense medium in the CGM of bright quasars \citep[e.g.][]{cantalupo2019large} (i.e., with a log-normal $\sigma>2$ if a log-normal density distribution is assumed).  \par
 In addition, our MOSFIRE spectroscopic observation allowed us to measure the detailed spectral profiles of H$\alpha$ emission in several regions within the slit and to compare them to the Ly$\alpha$ profiles. In particular, we observed in several regions a complex H$\alpha$ line profile with at least two separated components. These components, not present in the Ly$\alpha$ spectrum, show a relatively narrow line width and no significant velocity shifts as a function of distance from the quasar. Their narrow line profiles suggest rather quiescent kinematics in the CGM of J0010 which seems incompatible with quasar outflows capable of escaping the potential well of the host halo, while the presence of multiple kinematic components along the same line of sight also disfavours a scenario of disk-like rotation in a massive halo ($>10^{12}$M$_{\odot}$).
 Our results pave the way to future surveys for intergalactic H$\alpha$ emission, e.g. using IR Integral-Field spectroscopy with VLT/ERIS or JWST, which have the potential to provide a high-spatial-resolution view and, at the same time, detailed kinematical properties of the CGM/IGM. These properties can then be compared to numerical models of fluorescent H$\alpha$ emission in a cosmological context without the current complications associated with the Ly$\alpha$ emission modeling. Such a comparison will give us the opportunity to constrain both the quasar halo mass and the origin of the cold CGM component in terms, e.g., of cosmological inflows and galactic outflows.

\section*{Acknowledgements}

This project has received funding from the European Research Council (ERC) under the European Union’s Horizon 2020 research and innovation programme (grant agreement No 864361, PI: S. Cantalupo). SC gratefully acknowledges additional support from the Swiss National Science Foundation grant PP00P2\_190092 and from Fondazione Cariplo. GP acknowledges support from the Netherlands Research School for Astronomy (NOVA). SG acknowledges support from the Swiss National Science Foundation grant P2EZP2\_199856.

\section*{Data Availability}
The data underlying this article will be shared on reasonable request to the corresponding author.
%%%%%%%%%%%%%%%%%%%%%%%%%%%%%%%%%%%%%%%%%%%%%%%%%%



\bibliographystyle{mnras}
\bibliography{mybib} % if your bibtex file is called example.bib

%%%%%%%%%%%%%%%%%%%%%%%%%%%%%%%%%%%%%%%%%%%%%%%%%%

%%%%%%%%%%%%%%%%% APPENDICES %%%%%%%%%%%%%%%%%%%%%

\newpage

\section*{Appendix}
 
\appendix

\section{Flux Extraction Apertures} 
\label{app:apertures}
\begin{figure*}
\centering
\begin{subfigure}{.4\textwidth}
  \centering 
  \includegraphics[width=\linewidth]{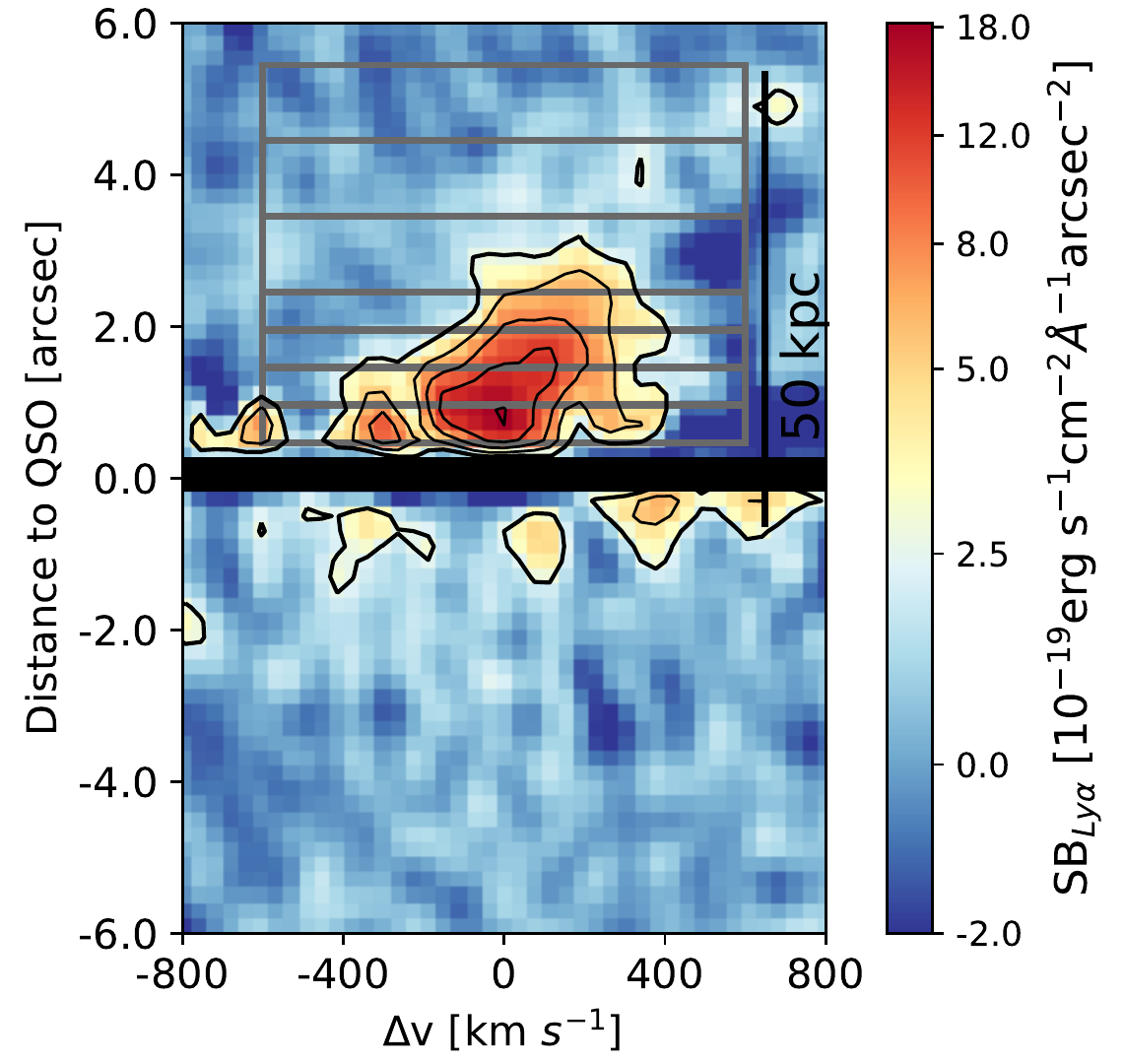}
\end{subfigure}%
\begin{subfigure}{.36\textwidth}
  \centering
  \includegraphics[width=\linewidth]{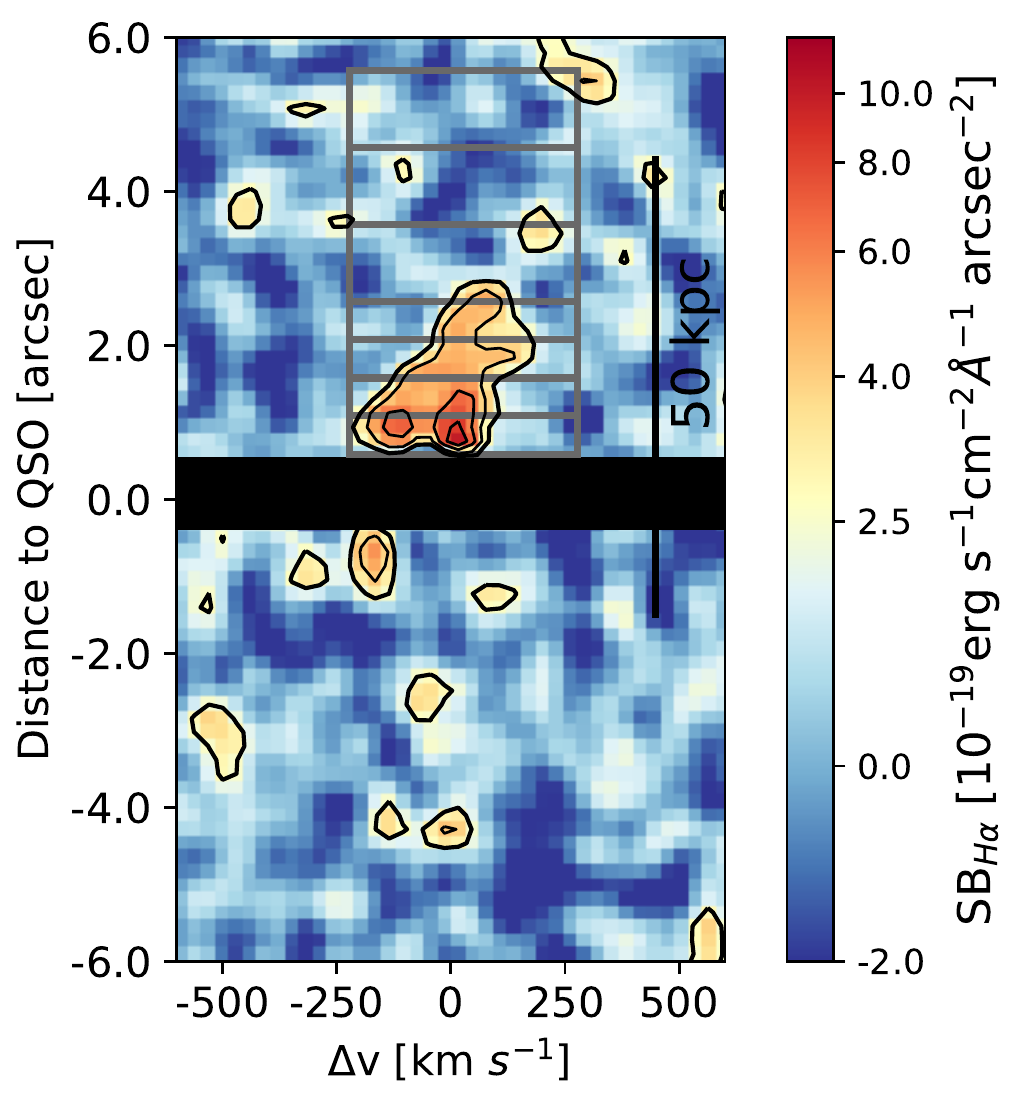}
\end{subfigure}
\caption[Extraction regions for the flux ratios in function of distance along the slit. ]{Presented here are the extraction regions for the flux ratios in function of distance along the slit. The spectral widths are 1200 km s$^{-1}$ and 500 km s$^{-1}$ for Ly$\alpha$ and H$\alpha$ respectively. }
\label{app:A1}
\end{figure*}

\begin{figure*}
\centering
\begin{subfigure}{.4\textwidth}
  \centering 
  \includegraphics[width=\linewidth]{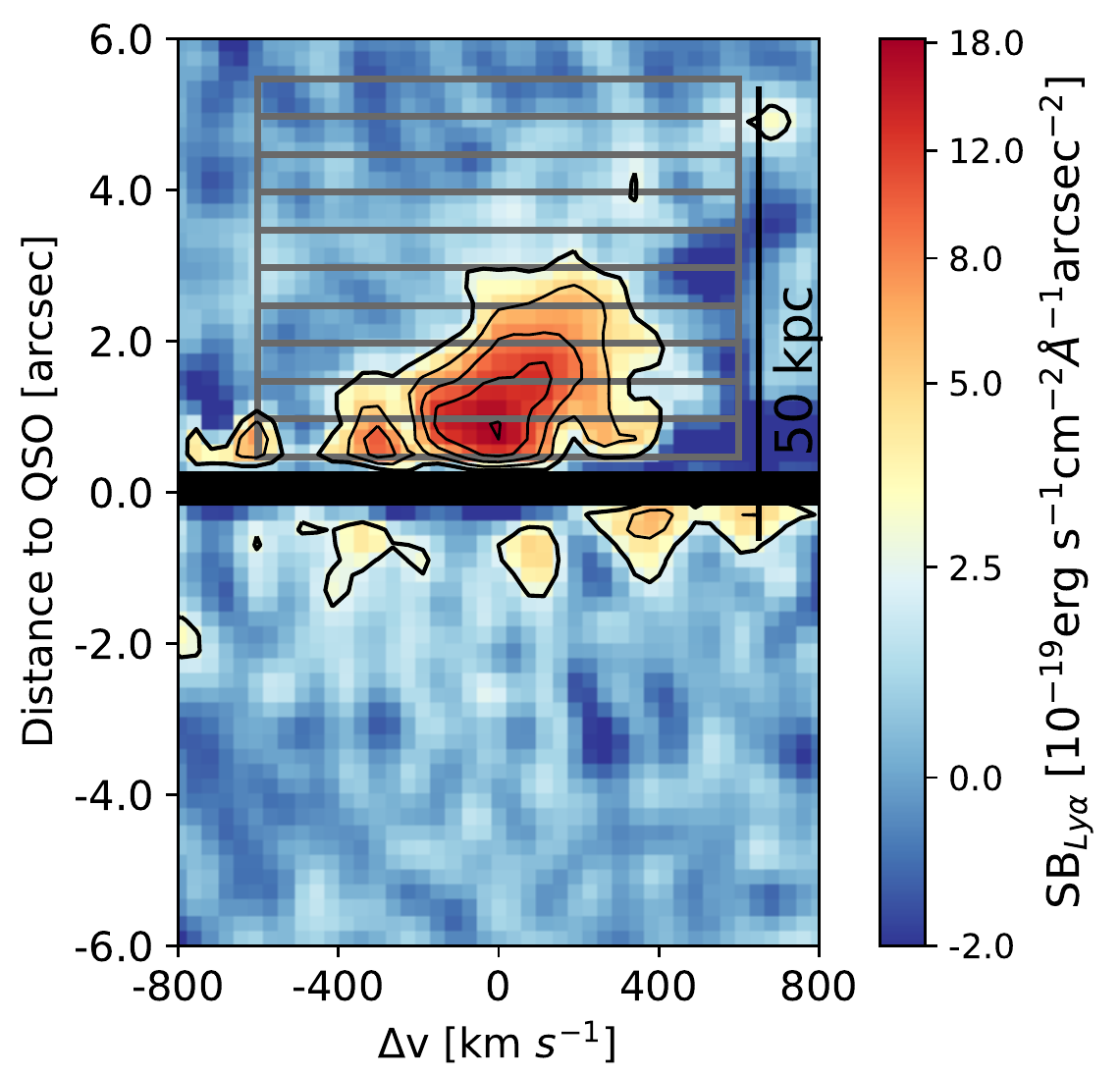}
\end{subfigure}%
\begin{subfigure}{.36\textwidth}
  \centering
  \includegraphics[width=\linewidth]{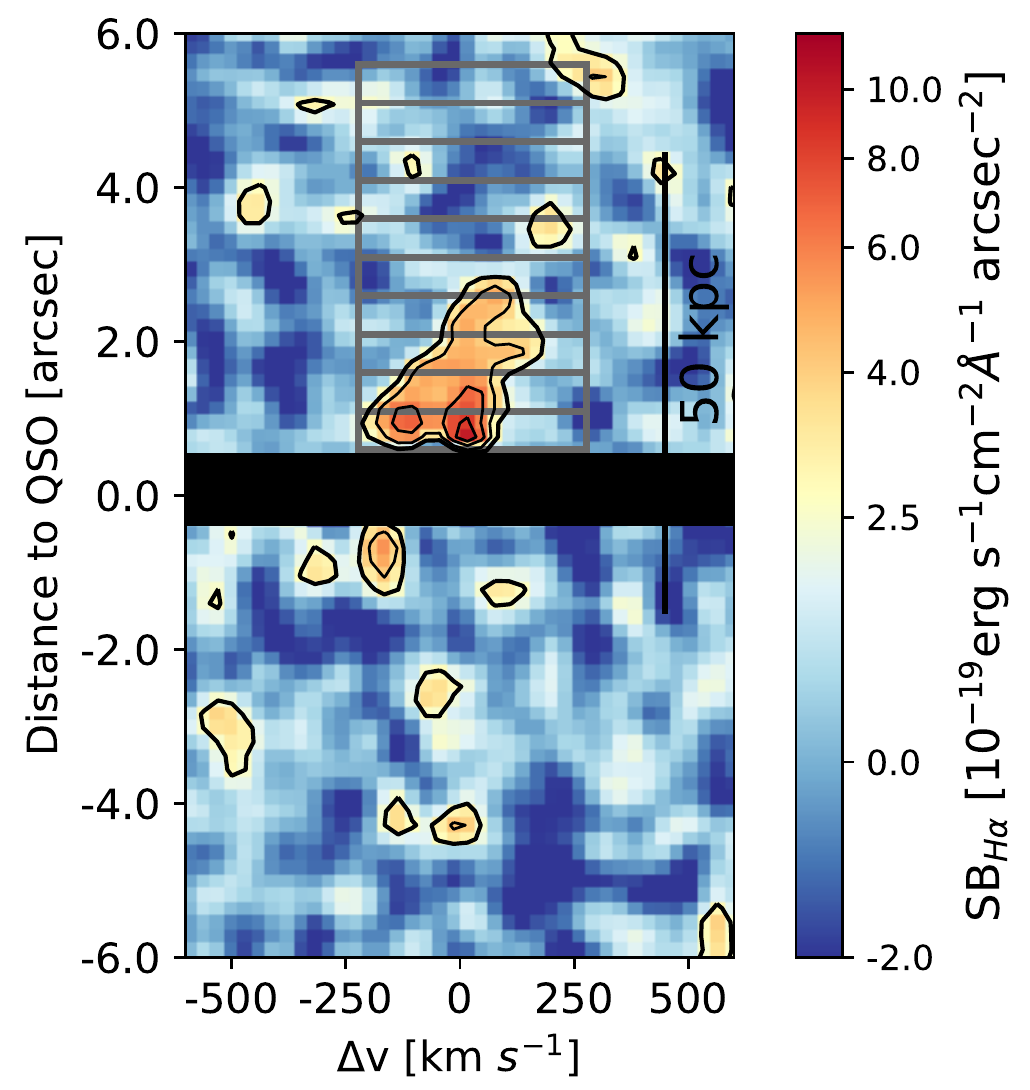}
\end{subfigure}
\caption[Extraction regions for the cumulative flux ratios]{Presented here are the extraction regions for the cumulative flux ratios. The spectral widths were chosen identical to the ones of the flux ratios along the slit, see Figure~\ref{app:A1}. }
\label{app:A2}
\end{figure*}

\begin{figure*}
\centering
\begin{subfigure}{.4\textwidth}
  \centering 
  \includegraphics[width=\linewidth]{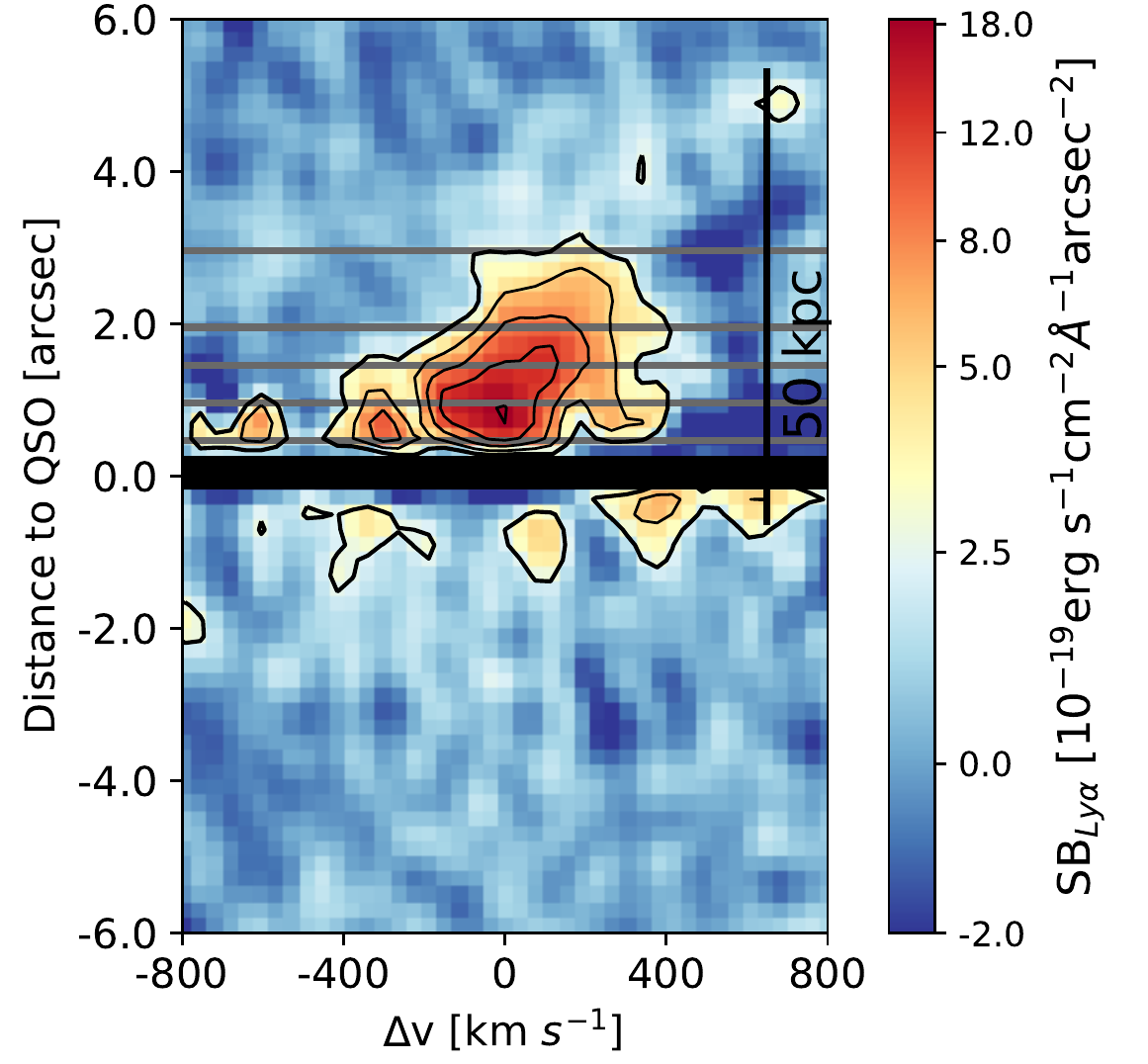}
\end{subfigure}%
\begin{subfigure}{.36\textwidth}
  \centering
  \includegraphics[width=\linewidth]{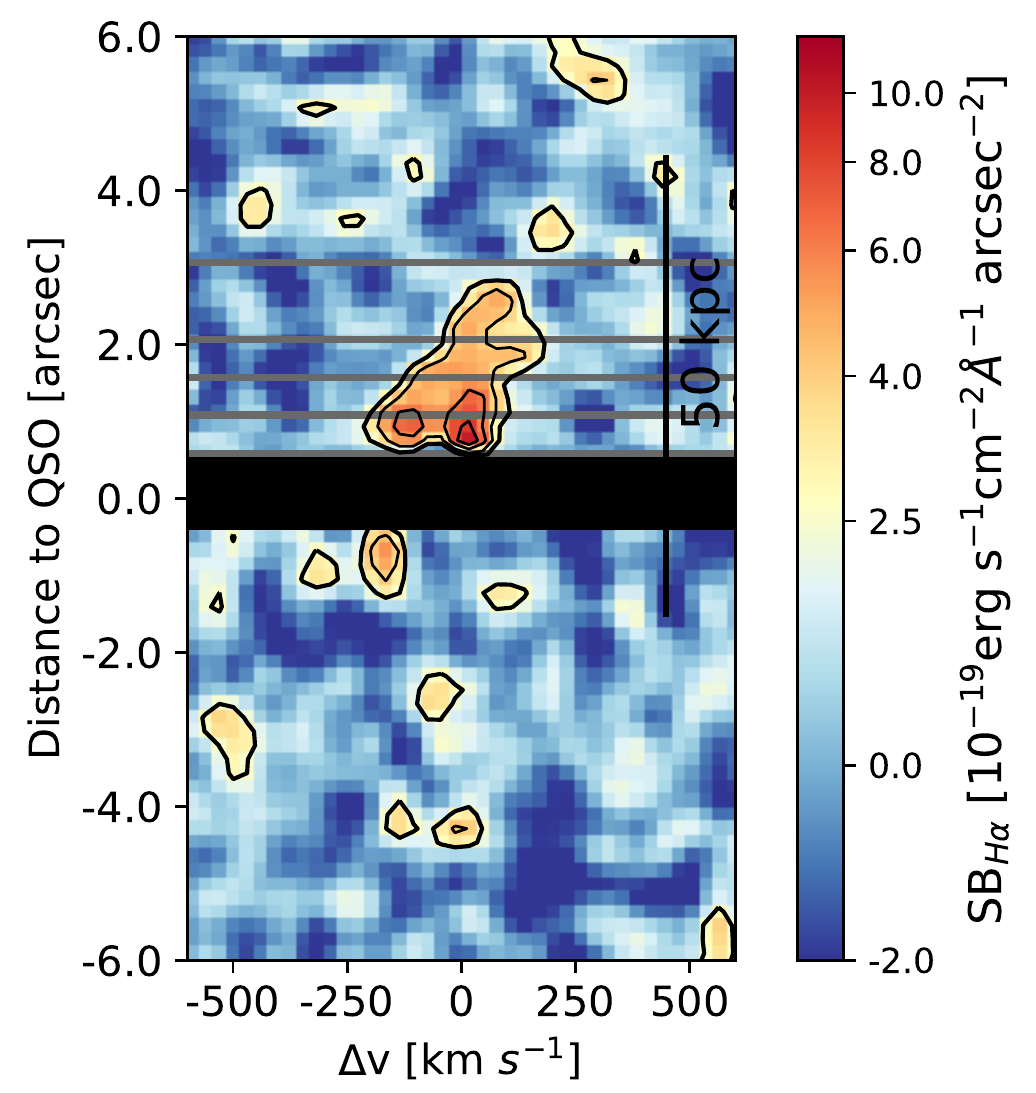}
\end{subfigure}
\caption[Extraction regions for the flux ratios and the comparison of the 1D spectra between Ly$\alpha$ and H$\alpha$]{Presented here are the extraction regions for the spectral projections of the flux densities along the spatial axis.  }
\label{app:A3}
\end{figure*}

Here we present the apertures that have been used to extract the Ly$\alpha$ and H$\alpha$ fluxes. In particular, we show the three different configurations used for our analysis of the flux ratios in function of distance to QSO, of the cumulative flux ratios and of the projection along the spatial axis, shown in Figure~\ref{app:A1}, Figure~\ref{app:A2} and Figure~\ref{app:A3} respectively. \\

\section{White Light Images}
\label{app:WLimages}
\begin{figure*} 
\centering
  \begin{subfigure}{0.3\textwidth}
    \centering  
    \includegraphics[width=\textwidth]{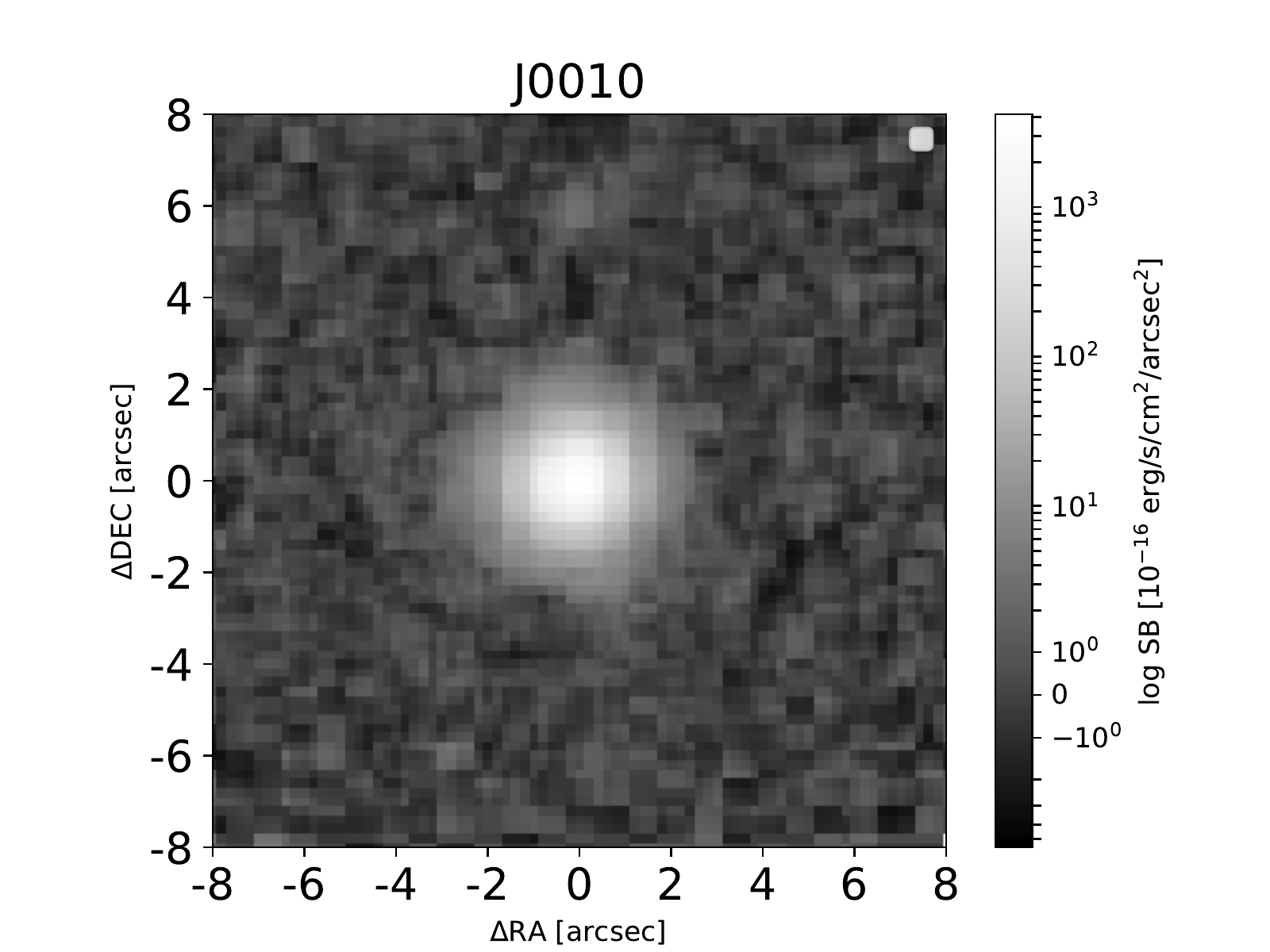}
  \end{subfigure}
  \hfill
  \begin{subfigure}{0.3\textwidth}
    \centering
    \includegraphics[width=\textwidth]{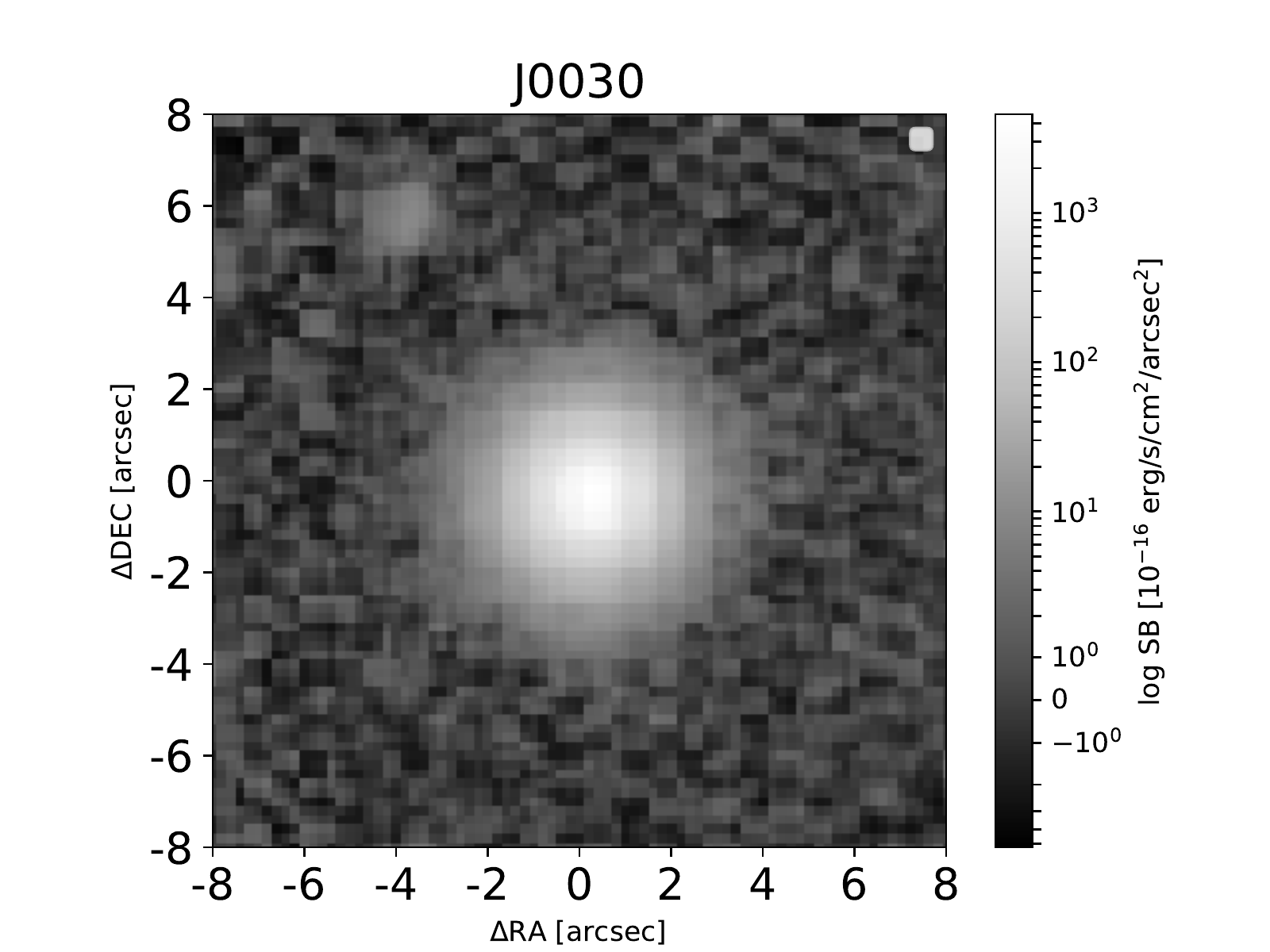}
  \end{subfigure}
  \hfill
  \begin{subfigure}{0.3\textwidth}
    \centering
    \includegraphics[width=\linewidth]{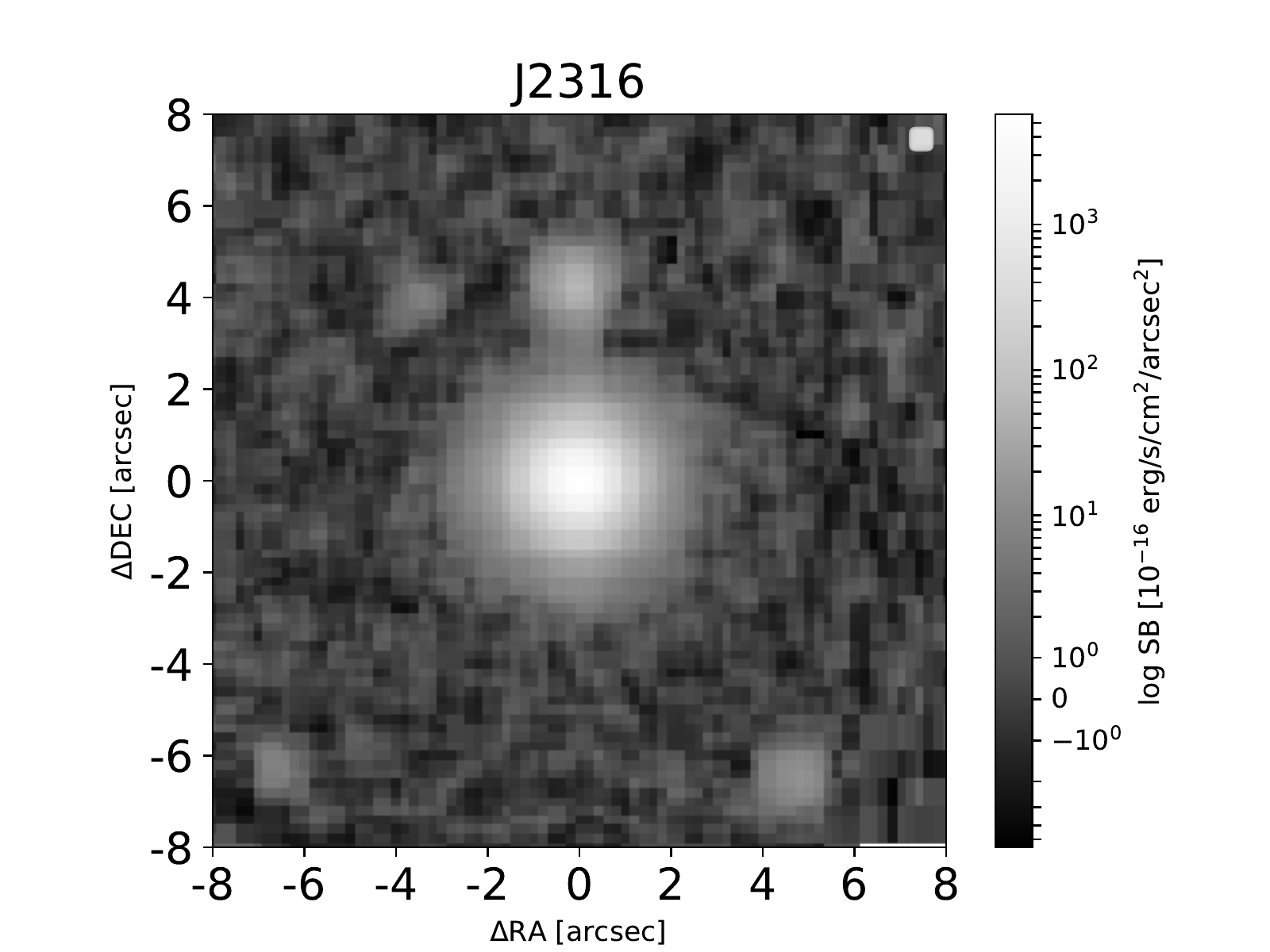}
  \end{subfigure}

  \begin{subfigure}{0.3\textwidth}
    \centering
    \includegraphics[width=\linewidth]{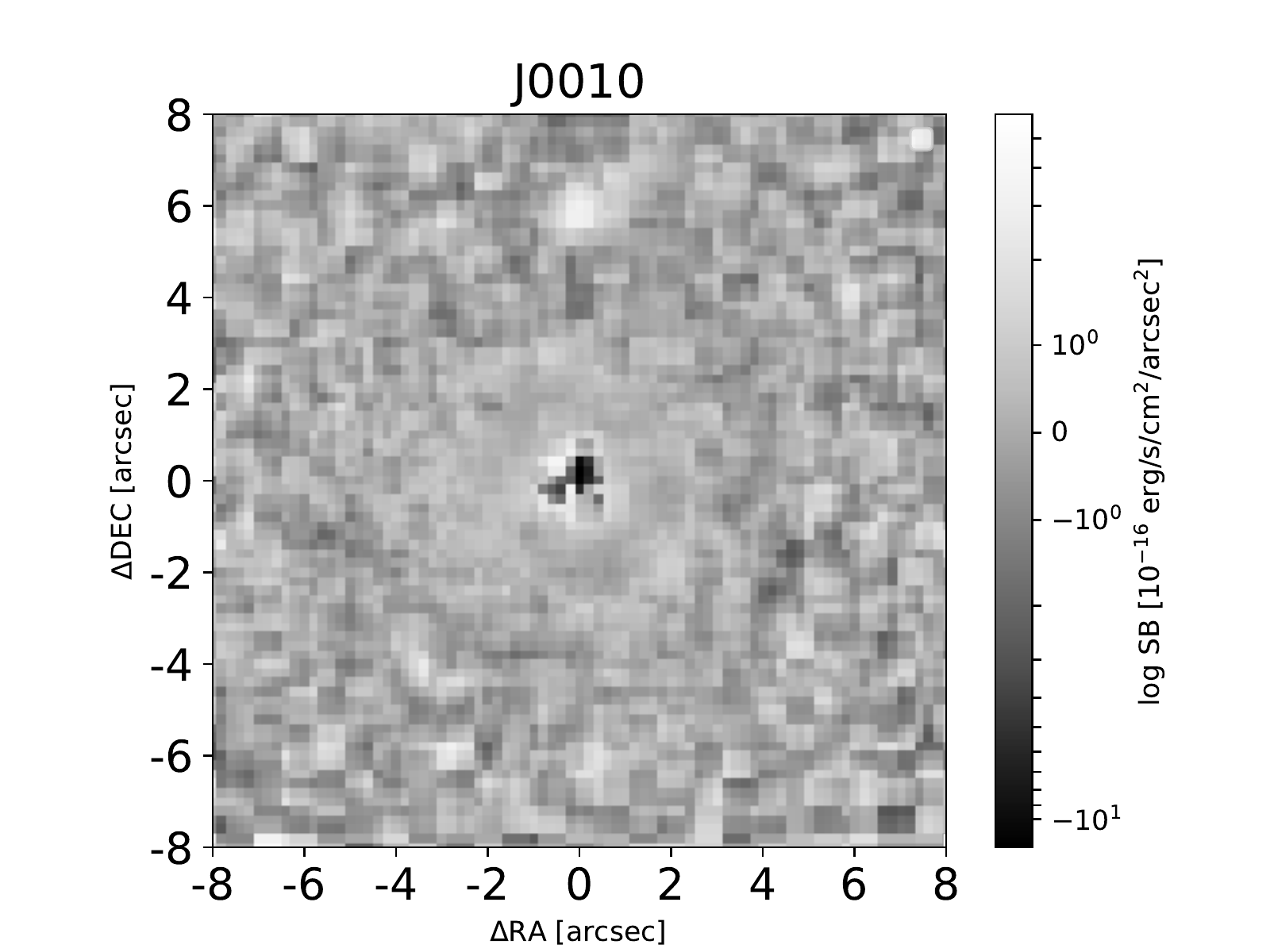}
  \end{subfigure}
  \hfill
  \begin{subfigure}{0.3\textwidth}
    \centering
    \includegraphics[width=\linewidth]{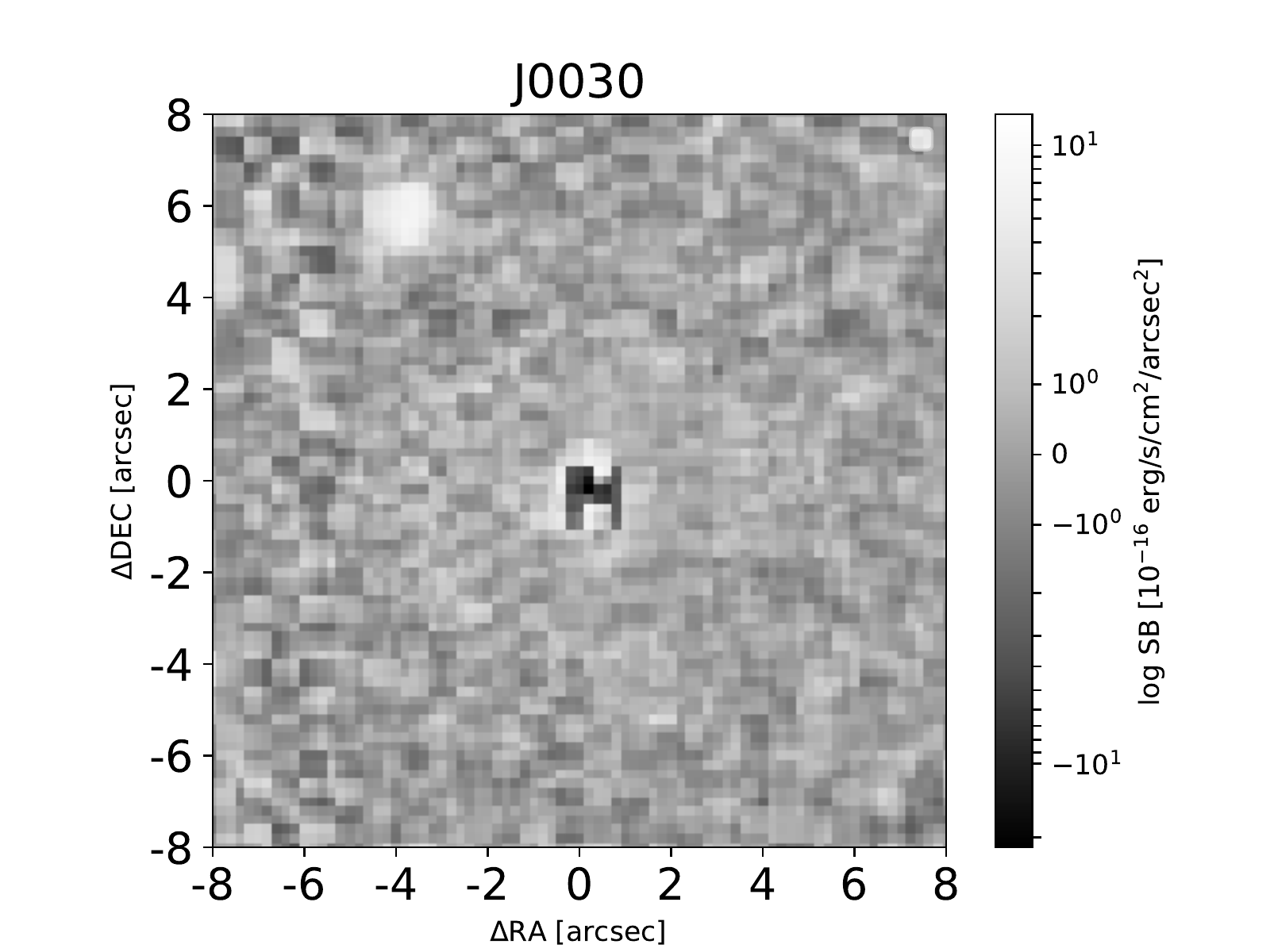}
  \end{subfigure}
  \hfill
  \begin{subfigure}{0.3\textwidth}
    \centering
    \includegraphics[width=\linewidth]{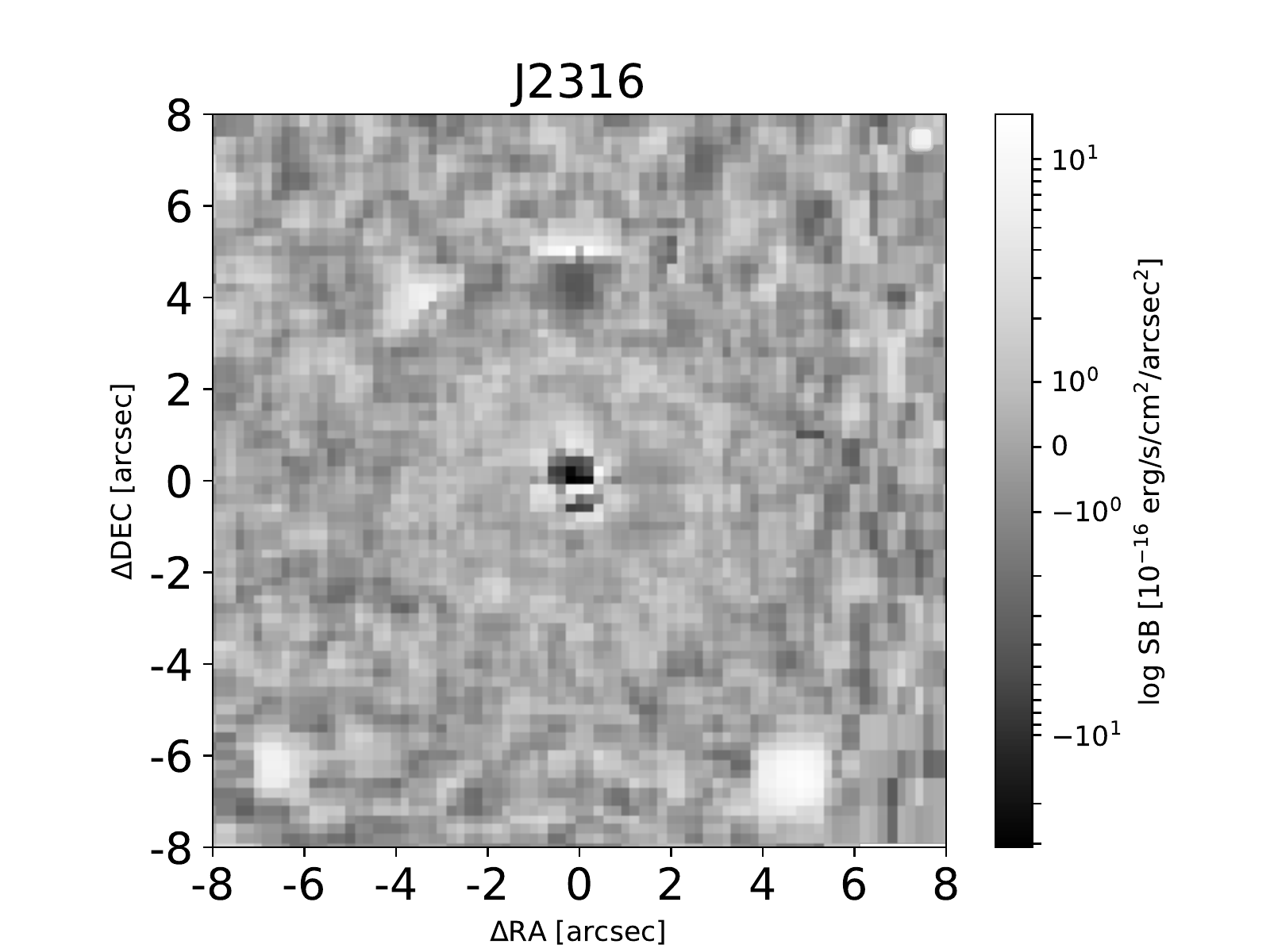}
  \end{subfigure}
 
\caption[White light images]{The upper row shows from left to right the WL images of the unsubtracted images of J0010, J0030 and J2316 respectively. The lower three panels are PSF subtracted and hence they show the continuum around the three QSO. Note that the images are plotted in log-scale. }
\label{app:WL}
\end{figure*}

In the following we present the white-light (WL) images of all three fields ones with and without PSF subtraction, in the upper and lower row of Figure~\ref{app:WL} respectively. We want to emphasize that no external continuum source can be associated to our detected H$\alpha$ emission. Hence we can conclude our detection to originate from the intrinsic nebula emission.

% Don't change these lines
\bsp	% typesetting comment
\label{lastpage}
\end{document}